\documentclass[12pt]{article}

\usepackage{graphicx}
\usepackage{latexsym}
\usepackage{caption2}
\usepackage{flafter}
\usepackage{cite}
\usepackage{amssymb,amsmath}                          
\usepackage[mathlines,displaymath]{lineno}
\usepackage{algorithmic}
\usepackage{algorithm}
\usepackage[all,2cell,dvips]{xy}

\title{Unifying Theories of Molecular, Community and Network Evolution}
\author{Carlos J. Meli\'an$^{1}$\footnote{To whom correspondence should be addressed. E-mail: melian@nceas.ucsb.edu,
phone: +1-805-892-2529, fax: +1-805-892-2510.}, David Alonso$^{2}$, Diego P. V\'azquez$^{3}$ and James Regetz$^{1}$\\
\\
\\$^1$National Center for Ecological Analysis and Synthesis,
\\University of California, 735 State St., Suite 300,
\\Santa Barbara, CA 93101, USA.
\\$^2$ Community and Conservation Ecology Group, University of Groningen
\\Haren, Groningen, The Netherlands.
\\$^3$Inst. Argentino de Investigaciones de las Zonas \'Aridas,
\\CONICET, CC 507, AR-5500, Mendoza, Argentina.}

\begin{document}

\maketitle
\baselineskip=8.75 mm

\linenumbers

\newpage

\begin{abstract}
{\small The origin of diversification and coexistence of genes and species have been traditionally studied in isolated biological levels. Ecological and evolutionary views have focused on the mechanisms that enable or constrain species coexistence, genetic variation and the genetics of speciation, but a unified theory linking those approaches is still missing. Here we introduce evolutionary graphs in the context of neutral theories of molecular evolution and biodiversity to provide a framework that simultaneously addresses speciation rate and joint genetic and species diversities. To illuminate this question we also study two models of evolution on graphs with fitness differences, which provide insights on how genetic and ecological dynamics drive the speed of diversification. Neutral evolution generates the highest speed of speciation, species richness (i.e. five times and twice as many species as compared to genetic and ecological graphs, respectively) and genetic--species diversity (i.e., twice as many as genetic and ecological graphs, respectively). Thus the speed of speciation, the genetic--species diversity and coexistence can differ dramatically depending on whether genetic factors versus ecological factors drive the evolution of the system. By linking molecular, sexual and trophic behavior at ecological and evolutionary scales, interacting graphs can illuminate the origin and evolution of diversity and organismal coexistence.}
\end{abstract}

\newpage

\section{Introduction}

One outstanding challenge in ecology and evolution is the development of an accurate and complete understanding of diversity across biological levels and spatial scales \cite{Strogatz:2001,Gould:2002,Levin:2006}. The neutral and nearly-neutral theories of molecular evolution \cite{Kimura:1968,King&Jukes:1969,Otha:1973} (hereafter $NTME$) were proposed to explain observations of high evolutionary rates and the maintenance of large amounts of molecular diversity within populations \cite{Kimura:1983,Nei:2005}. Similarly, the neutral theory of biodiversity (hereafter $NTB$) promises to contribute to our understanding of how species diversity is maintained in ecological systems \cite{Hubbell2:2001,Chave1:2004,Alonso:2006}. Both theories share the same framework \cite{Watterson:1974,Hu&He&Hubbell:2006}. Furthermore, both theories offer a baseline from which to extend theories of evolution \cite{Lynch:2007,Bernardi:2007} and to test the role of frequency and density dependent selection on the generation, evolution and maintenance of diversity at genetic and ecological levels, respectively \cite{Nei:2005,Gravel:2006,Levin:2006,Johnson&Stinchcombe:2007}.

Neutral models at the molecular level have considered mutation rate ($\mu$), random fluctuations of alleles (i.e., genetic drift), and molecular constraints on producing fertile offspring (i.e., the genetic similarity value $q^{ij}$ between any pair of individuals $i$ and $j$ must be higher than $q^{min}$) as mechanisms of speciation in populations with $J$ individuals \cite{Coyne:1992,Higgs&Derrida:1992,Gavrilets2:2004}. The neutral theory of biodiversity introduces the implicit speciation rate at the individual level ($\nu$) where species fluctuate randomly (i.e., ecological drift) and all individuals are equivalent (i.e., neutral competitive interactions) \cite{Moran:1962,Hubbell2:2001}. Speciation is crucial to the neutral biodiversity theory (without it diversity cannot be maintained), yet the speciation parameter is simply assumed and has no basis in biological processes. To integrate genetic and ecological neutral theories, we need to link the speciation rate ($\nu$) with explicit mechanisms of speciation from neutral molecular theories \cite{Gavrilets2:2004}.

Despite the striking parallels between neutral theories in population genetics and community ecology, the speed of speciation and diversity patterns at genetic and community levels have rarely been studied simultaneously \cite{Antonovics:1976,Vellend:2005,Whithmanetal:2006,Johnson&Stinchcombe:2007,Lankau&Strauss:2007}. This raises important questions. For example, let us consider a population with $J$ reproductive compatible individuals. This defines a completely connected graph of size $J$ $\times$ $J$. Given this initial graph in a population, does neutral evolution at molecular and ecological levels speed up speciation and increase genetic--species diversity? If frequency and density dependence effects at genetic and ecological levels are important, how can we discern the speed of speciation and genetic--species diversity under neutral or natural selection scenarios? Thus, do genetic or ecological level drive the speed of speciation, genetic--species diversity and coexistence? \cite{Hutchinson:1959,Felsenstein:1981,Bush:1993,Gavrilets1:2000}.

In order to answer those questions we need a framework that allows us to study the molecular and ecological levels simultaneously. This framework represents an ambitious research programme -- much more than can be accomplished in a single paper. Our goals here are more limited. First, we introduce evolutionary graphs \cite{Liebermanetal:2005} in the context of neutral theories of molecular evolution \cite{Kimura:1968,King&Jukes:1969,Higgs&Derrida:1992} and biodiversity \cite{Hubbell1:1979,Hubbell2:2001} which suggest a promising new way to provide a general account of how neutral, positive and negative density and frequency dependent selection affect the speed of diversification and genetic--species diversity. Second, we introduce genetic and ecological graphs where the genotype-phenotype of each individual are represented as one to one or are decoupled by the specific behavior and phenotypic plasticity of each individual, respectively. Note that in addition to the graph of reproductive individuals, we need to specify a new graph that captures the effect of the phenotypic plasticity in the system.

Evolutionary neutral graphs in the context of two mutualistically interacting populations are defined as follows. Consider two randomly mating populations of size $J_{R}$ and $J_{P}$ where each individual has an infinitely large genome sequence subject to random neutral mutations. The initial genetic similarity values between each pair of individuals ($q^{ij}_{R}$ and $q^{ij}_{P}$) in the matrices $Q_{R} = [q^{ij}_{R}]$ and  $Q_{P} = [q^{ij}_{P}]$ are equal to 1 and mutation rates $\mu_{R}$ and $\mu_{P}$ are equal among all individuals (Fig. 1a). At each time step, an individual of each population is chosen for death (Fig 1b). Two individuals are chosen for reproduction. Individuals have the same probability $1/J_R$ ($1/J_P$) to be chosen for reproduction or for death (Fig 1c). The offspring of these two individuals replaced the dead individual. The newborns in $R$ and $P$ can be the consequence of sexual reproduction without a mutualistic interaction (i.e., facultative mutualism given by $\omega$ $>$ 0) or a consequence of a mutualistic interaction with individual effectiveness $m$ between the first two chosen individuals for reproduction in community $R$ and $P$ (i.e., given by 1 - $\omega$) (Fig. 1d). 

All these elements allow us to develop models of evolution on genetic and ecological graphs with the following additions to the neutral model (Fig. 2): (1) fitness differences within each species according to the number of genetically related mating partners (i.e., genetic graphs), or to the number of trophic links with individuals in the second community (i.e., ecological graphs); (2) density dependence across species, thus rare species have higher probabilities of reproduction, and (3) contrary to genetic graphs, where the offspring can inherit the high connectance from its parents increasing its reproductive probability, all offspring in the ecological graph start with the same reproductive probability. Let us consider first the genotype-phenotype map as one to one. In this ``genotype--fitness speciation model'' (hereafter $GF$) reproductive probabilities are set according to the total number of genetically related individuals each individual $i$ can mate with, so that we take into account explicit fitness differences within each species (Fig. 2b). The genetic level, assuming that all the traits involved in sexual selection are under genetic control, determines the evolution of the system based on the genetic similarity among individuals.

There is empirical evidence for the effect of ecological interactions mediated by phenotypic traits on the evolution of diversity \cite{Janzen:1970,Connell:1978,Leigh:1999, McKinnonetal:2004,Nee:2005}, but they have so far been missing in neutral theories. Let us consider that the phenotype is not simply the product of the genotype, but that it is influenced by the interactions with individuals of a second community (i.e., second trophic level). In the ``phenotype--fitness speciation model'' (hereafter $PF$) we still have the genetic similarity constraint on having fertile offspring, but the role of ecological interactions is represented as a varying reproductive probability for each individual according to its specific behavior, development and phenotypic plasticity \cite{Jablonka&Lamb:2005}. Those phenotypic traits, not linked to the total number of genetically related matings with individuals of the same species, are given by the evolution of the number of trophic links with individuals of the second community. Thus, the reproductive probability of each individual increases with the number of trophic links with individuals of the second community, but it is independent of the number of potential genetically related matings with individuals of the same species (Fig. 2c).

We show that the neutral scenario, which is represented by a diverse genetic pool of parents in the context of decoupled evolving mating and trophic graphs, triggers the highest speed of speciation and highest levels of genetic--species diversity and coexistence. We also show that ecological graphs, whose reproduction is determined by specific behavior or phenotypic plasticity and not by the total number of genetically related matings, allow higher speciation rate and coexistence than mating graphs. Linking neutral theories at the molecular and ecological levels using evolving graphs promise to contribute to our understanding of contemporary diversity at multiple levels \cite{Davidson&Erwin:2006}. As we will demonstrate, it represents a powerful starting point to: 1) understand the speed of speciation and the relationship between genetic and species diversity by using genetic and ecological graphs \cite{Higgs&Derrida:1992,Gavrilets2:2004,Liebermanetal:2005}, and 2) understand the consequences of niche and neutral dynamics as a continuum that is based on ecological interactions among individuals \cite{McKane1:2004,Wootton3:2005,Gravel:2006}

\section{Results}

First, not surprisingly, mating and trophic number of links at the individual level are not correlated during the evolution of the system under the neutral and the phenotype fitness scenarios (Fig. 3a and Fig. 3b, respectively). The distribution of Spearman's rank coefficient values is close to a uniform distribution in both cases (Fig. 3a and 3b represent the distribution for community $J_R$). The distribution of the Spearman's values for the genotype fitness model is highly skewed with approximately $80\%$ of $p$-values $<$ $0.01$, suggesting that mating and trophic degree are in most cases correlated. Fig. 3d and 3e represent the evolution of individual mating and trophic degree in the genotype fitness model as a function of the individual rank (i.e., from the most (left) to the least (right) connected individual).

The neutral unified model generates on average twice and five as many speciation events (i.e., $188\pm10$) as the phenotype (i.e., $80\pm1$) and the genotype fitness models (i.e., $34\pm3$) respectively (Fig. 4a, results for the community $J_{P}$ not shown but with the same parameter values they are qualitatively the same). Similarly, waiting time to speciation or the number of generations to the first speciation event is on average twice and five times as small in the neutral case ($170\pm3$) as in the phenotype ($440\pm12$) and the genotype fitness scenarios, respectively ($924\pm25$) (Fig. 4a, see appendix for a detailed description of the sampling of the transients and the steady state). At stationary state (approx. 1000 generations, see Fig. 5) speciation events scale with the number of generations for all the three models ($r^2$ = 0.99) with the scaling exponent $\gamma$ = 0.97 (neutral), 1.03 (phenotype fitness) and 1.31 (genotype fitness), red lines in Fig. 4a. 

Note that we have used the same three input parameters in the three models explored. Mutation rate, with $\mu_{R}$ = $\mu_{P}$ = $\mu$, the minimum genetic similarity value $q^{min}_{R}$ = $q^{min}_{P}$ =  $q^{min}$ and the individual mutualistic effectiveness $m_{ij} = m_{ji} = m = 1$ assuming a fully symmetric case for all the individual interactions in the context of obligate mutualism (i.e., $\omega$ = 0) (see Methods). Thus, the speed of speciation rate is driven by the specific reproductive transition probabilities at individual level. This result remains similar after relaxing the assumptions of effectiveness and facultative or obligate mutualism. Does the distribution of the number of generations to speciation differ among the models? All the nontransformed distributions were highly skewed (skewness index $>$ 2), and significantly different from a normal distribution (Fig. 4b $Lilliefors's$ test, all P $<$ 0.001 with means of $47$, $258$, and $115$ for the neutral, the genotype and the phenotype scenario, respectively). The distribution of the number of generations to speciation differ significantly between the neutral and the genotype/phenotype models ($Kolmogorov-Smirnov$ test, P $<$ 0.0001), but the genotype and the phenotype fitness scenarios do not differ significantly ($Kolmogorov-Smirnov$ test, P $>$ 0.1). 

The neutral model generates on average twice as many genetic and species diversity as the phenotype and the genotype fitness scenarios (Fig. 5a using eq. (2) in Methods, and 5b, using species diversity $S_{e}$ as $\frac{1}{\sum_{i}^{S_{e}} {p_i}^2}$ , where $p_{i}$ is the relative abundance of species $i$). As in the speed of speciation, the neutral scenario predicts twice and five as many number of coexisting species as the phenotype and the genotype fitness model, respectively (Fig. 5c). Genetic diversity (Fig. 5a), species diversity (Fig. 5b) and species richness (Fig. 5c) distributions for all the models differ from a normal distribution ($Lilliefors'$ test, P $<$ 0.001) despite their strong differences in skewness. The neutral case predicts highly symmetric distributions, all skewness indices between -0.15 and 0.08, while the phenotype and the genotype model predict skewness indices between 0.87 and 1.44 and $>$ 2, respectively. Genetic--species diversity and species richness distributions differ significantly among all the models ($Kolmogorov-Smirnov$ test, P $<$ 0.0001). 

In summary, the diverse genetic pool underlying our unified neutral scenario in the context of the uncorrelated mating and trophic graphs triggers the highest speed of speciation with consequences to the genetic-species diversity, coexistence and species richness. Note, however, that the species diversity values with the explicit mechanisms of speciation are lower than the values from the biodiversity number $\theta_{b}$ in the neutral theory of biodiversity. These results remain qualitatively similar for the range of parameter combinations explored (see appendix).

\section{Summary and Discussion}

The present study is an attempt to unify the speed of speciation with the evolution of diversity at genetic and ecological levels. We create a bridge between the neutral theory of molecular evolution \cite{Kimura:1968,King&Jukes:1969} and the neutral theory of biodiversity \cite{Hubbell2:2001} using mating and ecological graphs in the context of explicit mechanisms of speciation \cite{Higgs&Derrida:1992,Gavrilets3:2003}. The unified neutral model predicts the highest speed of speciation, number of coexisting species (i.e., five and twice as many as genetic and ecological networks, respectively), and genetic--species diversity (i.e., twice as many as genetic and ecological networks), but diversity values are lower than the neutral biodiversity theory with implicit speciation. This result is not surprising. Genetic variation maintained in non random mating is to same extent cryptic since the heterozygote diversity is less than from a random mating population. However we show how the speed of speciation and genetic-species diversity are closely controlled by the dominant graph (i.e., genetic or ecological) at each level during the evolution of the system.

Note that we have explored only a few scenarios (see appendix). Despite that the effect of the genetic regulatory \cite{Morata&Laurence:1977,Lewis:1978,Davidson&Erwin:2006} and ecological interactions \cite{Janzen:1970,Connell:1978,Leigh:1999,Nee:2005} on the evolution of diversity is widely recognized, they have so far been missing in neutral theories. Here we show that the decoupling of phenotypic (i.e., based on ecological interactions) from genotypic evolution (i.e., based on mating--genetic interactions) speeds up diversification and approaches to the neutral scenario. Evolutionary graphs have many fascinating extensions. For example, does frequency dependence selection at genetic level trigger higher speed of speciation and diversity than the neutral scenario? how do gene regulatory and mating graphs interact to jump from micro to macroevolution?

How does sexual reproduction affect evolution on graphs? Here we show that constraining fitness according to the total number of potential matings or trophic interactions per individual (i.e., the genotype or phenotypic fitness model, respectively), which implies most connectivity clustered in few individuals, are a potent selection amplifier \cite{Liebermanetal:2005}, and suppresses speciation rate, genetic--species diversity and species richness for all the range of mutation rates and the minimum similarity values explored. This cost to diversification by common parentage factor scaling up from individuals to genetic and ecological graphs adds an additional constraint to the cost of being excessively abundant or rare \cite{Gavrilets2:2004} and the metabolic cost \cite{Allenetal:2006}, thus how does the evolution of metabolic rate interact with sexual and ecological graphs to enhance or constrain diversity at multiple biological levels and spatial scales?

Most models of sympatric speciation rely on (1) intraspecific competition to drive divergence and reproductive isolation without specifying the niche or neutral nature of the interactions \cite{HTY:1999,KK:1999,DD:1999}, and (2) ecological dynamics that focus on the waiting time to the first speciation \cite{Gavrilets3:2003,Bolnick:2006}. On the other hand, neutral theory in community ecology studies patterns at the community level based on implicit modes of speciation with incipient species abundance $J^{s}$ $\geq$ $1$ \cite{Hubbell2:2001,Hubbell&Lake:2003,Gavrilets2:2004,Hu&He&Hubbell:2006,Bolnick:2006}. Here, despite the importance of explicit space, local adaptation and explicit prezygotic/postzygotic isolating factors to determine the mode and speed of speciation \cite{Malecot:1970,Coyne:1992,Manzo&Peliti:1994,Gavrilets2:2004,Nei:2005,Vellend:2005}, we link the first speciation event with the speed of speciation (i.e., mutation and fission modes of sympatric speciation), the number of coexisting species and the genetic--species diversity in a unified framework. Note that the speed of speciation for all the parameter combinations and models explored is extremely high. On average it is $47$, $115$, and $258$ generations to speciation, for the neutral, the phenotype and the genotype scenarios, respectively (see however \cite{Hendryetal:2007}). If we assume a linear extrapolation from $J_{R}$ ($J_{P}$) = $10^{3}$ to $10^{5}$ inds., $\mu$ = $10^{-4}$ to $10^{-6}$, and $q^{min}$ = $0.9$ ($Q_{R}$ ($Q_{P}$) $\sim$  ${Q_{R}}^{*}$ (${Q_{P}}^{*}$) $\sim$ 0.7, see eq. 1), then the number of generations to speciation approaches to $4.7 \times 10^{3}$, $11.5 \times 10^{3}$, and $25.8 \times 10^{3}$, which are close to the observed values in more realistic sympatric speciation models (i.e., less than $2 \times 10^{4}$ \cite{Gavriletsetal:2007} and $5 \times 10^{4}$ \cite{Gavrilets&Vose:2007} generations). 

Studies on food webs assume species level approaches despite the intrinsic variability in individuals \cite{Bolnicketal:2002}. In the genotypic and phenotypic fitness models only a few individuals within each population (i.e. ``hubs'') drive reproductive rate in the context of symmetric effectiveness of ecological interactions. The expected outcome by coupling fitness with competitive and trophic asymmetry at ecological level would inevitably decrease species richness, coexistence and diversity by decreasing persistence probabilities of individuals with lower fitness. This suggest that individual variability, driven by the degree of symmetry between each pair of interacting individuals and the effectiveness of each interaction, can dramatically alter the speed of speciation, genetic-species diversity, coexistence and the structure of food webs. Note that ``hubs'' in networks are common but their role in inhibiting or expressing speciation and diversification at different biological levels is still unknown \cite{Strogatz:2001}. The need of food web data at individual level is then crucial to determine how interacting graphs at genetic and ecological levels generate the patterns of diversity and coexistence of food webs. For example, do ecological interactions depend of species or individual traits? are ecological interactions governed by a few number of individuals within each population? does neutral evolution predict the complexity and the structure of food webs?

Rapid accumulation of empirical results from different biological levels suggests that ecological and evolutionary theory are undergoing a change \cite{Howard&Berlocher:1998,Hubbell2:2001,Jablonka&Lamb:2005}. The need to test models from first principles is now widely recognized \cite{Gavrilets2:2004,Levin:2006,Whithmanetal:2006,Lynch:2007,Johnson&Stinchcombe:2007}. Here we present a unified neutral model of evolution with mutation, mating with random mixing of genes, genetic--ecological drift and neutral interactions as the driving forces of diversity at multiple levels in three different scenarios. Promisingly, the huge amount of data collected and meticulously cataloged at each biological level can be used to test neutral models from first principles in a general niche--neutral continuum multilevel framework \cite{Hubbell2:2001,McGill:2003,Wootton3:2005,Nee:2005,Gravel:2006,Levin:2006,Johnson&Stinchcombe:2007}.

\section{Methods: Unifying  Molecular and Ecological Evolution}

We first describe the Higgs and Derrida model of neutral molecular evolution \cite{Higgs&Derrida:1992} with explicit mechanisms of sympatric speciation \cite{Bolnick&Fitzpatrick:2007}. Second, we describe the Hubbell's neutral model of biodiversity \cite{Hubbell2:2001} with implicit speciation. Third, we highlight their similarities and link those models in the context of two initial populations that give rise to two mutualistically interacting communities. Finally we show how this framework allow us to compare the speed of speciation and the genetic--species diversity between the neutral scenario and two models of evolution at genotypic and phenotypic levels \cite{Haldane:1957} using genetic and ecological graphs \cite{Liebermanetal:2005}, respectively.

\subsection{Neutral Molecular Evolution Model}

Our starting point is a basic stochastic model for species formation by Higgs and Derrida (1992). This model contains three nonadaptive evolutionary forces in the sense that they are not a function of the fitness properties of the individuals: 1) neutral mutation rate ($\mu$) in diploid and hermaphroditic individuals with equal and independent changes across any locus in a infinite genome size \cite{Kimura&Crow:1964}; 2) mating with neutral mixing of genes from an hermaphroditic or two nonidentical parents and 3) genetic drift, which ensures that gene frequencies will deviate slightly from generation to generation independent of other forces \cite{Higgs&Derrida:1992,Lynch:2007} (see section A1 in the Appendix). 

Consider one initial completely connected and randomly mating population of size $J$, where individuals have the same genetic sequence. The initial genetic similarity values between any pair of individuals ($q^{ij}$) in the genetic similarity matrix $Q = [q^{ij}]$ are equal to 1. At each time step, an individual is chosen for death and two individuals are chosen for reproduction. Individuals have the same probability to be chosen for reproduction or for death ($1/J$). The viability of the offspring is constrained by $q^{min}$, defined as the minimum genetic similarity value for postzygotic reproductive isolation ($RI$) two individuals $i$ and $j$ must satisfy for the development of fertile offspring \cite{Mayr:1970,Coyne:1992,Higgs&Derrida:1992,Wu:2001,Gavrilets2:2004}. Thus, in a randomly mating population this minimum value works as a filter generating viable offspring if and only if $q^{ij}$ $>$ $q^{min}$. The offspring of these two individuals replace the individual that died.
 
In this model, if the mutation rate is low ($\mu$ $<<$ 1), then the mean similarity value for $Q$ has a solution \cite{Kimura&Crow:1964,Higgs&Derrida:1992}

\begin{eqnarray}
{Q}^{*} = \frac{1}{4 J \mu + 1} \label{solution similarity}\\ \nonumber
\end{eqnarray}

where $4 J \mu$ = $\theta_{m}$. The mean value arises because of a
balance between mutations (which decrease the average similarity value,
$\langle Q \rangle$) and the common parentage factor which is given 
by the probability that two individuals have a common ancestor (which
increase $\langle Q \rangle$). Similarly, the probability that two
individuals do not have a common ancestor at stationarity is given by

\begin{eqnarray}
1 - {Q}^{*} =  \frac{\theta_{m}}{\theta_{m} + 1}\\ \nonumber
\end{eqnarray}

Results from eqs. (1) and (2) are similar to the probability that one individual is homozygous or heterozygous for one single locus under the neutral molecular theory, respectively 
\cite{Kimura&Crow:1964}. If $q^{min}$ $<$ ${Q}^{*}$ we will have always one species with size $J$. Eq. (2) represents a measure of genetic diversity in the population $J$.

Interestingly, if $q^{min}$ $>$ ${Q}^{*}$, then the initial population $J$ is greatly perturbed by the cutoff, which implies that the genetic similarity matrix (${Q}$) can never reach its equilibrium
state. As a consequence, the initial population splits and speciation happens with the species fluctuating in the system according to demographic stochasticity \cite{Higgs&Derrida:1992} (see section A1 in the Appendix). 

\subsection{Neutral Theory of Biodiversity}

The neutral theory of biodiversity considers species instead of alleles and introduces the implicit speciation rate at the individual level ($\nu$). The standard evolutionary metacommunity model assumes that at each time step one individual is chosen to die with probability $1/J$ and is replaced by the newborn. With probability $1 - \nu$, this new individual is of the same species as its parent; with probability $\nu$, it is an entirely new species \cite{Moran:1962,Hubbell2:2001}. 

In this context, the probability that individual $i$ and $j$ in population $J$ chosen at random will be of the same species is 
\begin{eqnarray}
{f}^{*} = \frac{1}{\theta_{b} + 1}\\ \nonumber
\end{eqnarray}
where the biodiversity number ($\theta_{b}$) is equal to $J \nu$. 

In this scenario ecological drift dominates. Each individual has a percapita probability to speciate at each reproduction event. This point mutation model of implicit speciation is an individual level process that leads to a proportional relationship between the speciation rate of each species in a community and its abundance \cite{Hubbell2:2001,Etiennetal:2007}.  

In summary, these two neutral models describe a zero--sum evolving
population of $J$ individuals with overlapping generations under
demographic stochasticity. The neutral molecular model starts with a
completely connected graph with mutations and mating with random
mixing of genes adding variation to the new individual with explicit
speciation if $q^{min}$ $>$ ${Q}^{*}$. The biodiversity model includes
the implicit speciation parameter $\nu$. In the next section we link the implicit speciation rate ($\nu$) with
explicit mechanisms such as mutation rate ($\mu$) and the minimum
similarity value for postzygotic reproductive isolation
($q^{min}$). Both neutral models are based on one initial population
that gives rise to one community. In the next sections we describe the
link between neutral molecular and biodiversity theory in the context
of two initial populations that will give rise to two interacting
communities.  

\subsection{Unified Neutral Model: Molecular and Ecological Evolution}

The dynamics of our first two-community model has stochastic birth and death as Hubbell's model, but considers mutation ($\mu$) and the minimum  similarity value ($q^{min}$) as explicit mechanisms of speciation at the molecular level \cite{Higgs&Derrida:1992} and the effectiveness of each mutualistic interaction at ecological level. We consider sexual diploid populations with overlapping generations and age independent birth and death rates. The individual interactions occur in a single homogeneous patch \cite{Hubbell2:2001,McKane1:2004}. The two populations can be thought of as hermaphroditic plants and dioecious pollinators, which respectively are labeled $R={1,\dots,J_R}$ and $P={1,\dots,J_P}$,  where $J_R$ and $J_P$ are the total number of individual plants and pollinators, respectively. The total number of individuals is $J_{m}$ = $J_R$ + $J_P$, which implies that all individuals are considered to be of reproductive age in the metacommunity.

The basic model has three input parameters (i.e., the mutation rate assuming $\mu_{R}$ = $\mu_{P}$ = $\mu$, the minimum genetic similarity value assuming $q^{min}_{R}$ = $q^{min}_{P}$ = $q^{min}$, and the effectiveness of each mutualistic ecological interaction, $m$, assuming a fully symmetric case for all the individuals interactions) and two
explicit biological levels: 1) genetic, represented as mutation
($\mu$), mating with random mixing of genes, genetic drift and the
$q^{min}$ as in the Higgs and Derrida model already described
\cite{Kimura&Crow:1964,Higgs&Derrida:1992,Lynch:2007}, and 2)
ecological level represented as ecological drift as in Hubbell's model
\cite{Hubbell2:2001} in the context of equal and symmetric
competitive and mutualistic ($m$) ability among all interacting
individuals. This is the simplest neutral scenario but the framework
allows extensions to more complicated ones.

The speed of speciation is then governed in each plant and pollinator
species by the mutation rate ($\mu$), $q^{min}$ for $q^{min}$ $>$
${Q}^{*}$, and the type of sexual reproduction (i.e., facultative or
obligate, mediated by the mutualistic effectiveness parameter, $m$). The rate of decay of genetic similarity of the newborn $j$ given the similarity between the parents ($k^{'}_{1}(j)$,$k^{'}_{2}(j)$) of the new individual $j$ and each individual $i$ already in the population is given by (see appendix):
\begin{equation}
q^{ji}  = \frac{e^{-4\mu} }{2} (q^{k^{'}_{1}(j) i} + q^{k^{'}_{2}(j) i}),\label{A3}
\end{equation}
\noindent where $k^{'}_{1}(j)$ can be the same than $k^{'}_{2}(j)$ in the hermaphroditic plant species. The time evolution of the plant and the pollinator species are governed by the generalized birth and death process with the probability of speciation in the hermaphroditic plant ($\nu^k_{R}$) and dioecious pollinator ($\nu^k_{P}$) species $k$ represented as:
 
\begin{equation}
\nu^{k}_{R} = \frac{2}{N^{k}_{R} (N^{k}_{R} + 1)} \sum_{k^{'}_{1} = 1}^{N^{k}_{R}} \sum_{k^{'}_{2} = k^{'}_{1}}^{N^{k}_{R}} P^{k^{'}_{1},k^{'}_{2}}_{R},
\end{equation}

\begin{equation}
\nu^{k}_{P} = \frac{2}{N^{k}_{P} (N^{k}_{P} - 1)} \sum_{k^{'}_{1} = 1}^{N^{k}_{P}} \sum_{k^{'}_{2} = k^{'}_{1} + 1}^{N^{k}_{P}} P^{k^{'}_{1},k^{'}_{2}}_{P},
\end{equation}
\vspace{0.2 in}

where $P^{k^{'}_{1},k^{'}_{2}}_{R}$ and $P^{k^{'}_{1},k^{'}_{2}}_{P}$ are defined as the probabilities to produce a new species from two randomly chosen individuals ($k^{'}_{1},k^{'}_{2}$) in the plant or pollinator species $k$, respectively:
\begin{equation}
{\Large P^{k^{'}_{1},k^{'}_{2}}_R = F\left[ \frac{\sum_{\substack{i = 1\\ i \neq j}}^{J_{R}} H(q_t - (q^{k^{'}_{1} i} + q^{k^{'}_{2} i}))}{J_R - 1} \right]}
\end{equation}
\begin{equation}
{\Large P^{k^{'}_{1},k^{'}_{2}}_P = F\left[ \frac{\sum_{\substack{i = 1\\ i \neq j}}^{J_{P}} H(q_t - (q^{k^{'}_{1} i} + q^{k^{'}_{2} i}))}{J_P - 1} \right]}
\end{equation}
\vspace{0.2 in}

where $q_t$ = $\frac{2 q^{min}}{e^{-4 \mu}}$, $F(x) = 1$ if $x=1$ and zero otherwise. $H(\alpha)$ is 
\begin{equation*}
H(\alpha) = \\
\begin{cases}
1 & \text{if $\alpha > 0$}\\
0 & \text{otherwise} 
\end{cases}
\end{equation*}

Note that we have two modes of speciation. Expressions above characterize a mutation mode of speciation. $P^{k^{'}_{1},k^{'}_{2}}_R$ either is 1 when speciation occurs and zero otherwise. When the offspring of two individuals is a new individual $i$ that cannot mate with any individual in the community (with $i$ $\neq$ $j$), we have an incipient species of size 1. However, death events may induce the formation of new species of larger sizes. When there is only one individual connecting two mating networks and this happens to die, a new species arises. This speciation process can be called a fission speciation mode.

In summary, our unified neutral model represents the stochastic evolution of two initial finite populations that give rise to two interacting communities (see Fig. 2a). Therefore, the interactions among individuals belonging to two different communities trigger the development of the ecological network as a consequence of the neutral dynamics at molecular and ecological levels. 

\subsection{Evolution on Graphs: The Genotype and the Phenotype Fitness Speciation Model}

Do genetic or ecological mechanisms drive the speed of speciation, genetic--species diversity and coexistence? To illuminate this question we describe two alternative models of evolution on graphs with explicit
individual fitness differences. Our goal is to compare them with our unified neutral model, introduced in the last section. Fitness is defined for each individual as the reproductive probability according to the genetic similarity (i.e., genotype fitness model) or ecological affinity (i.e., phenotype fitness model) with other individuals in the same population or with individuals of the other community, respectively, but at the same time we keep neutral mutations at the genetic level symmetric. Apart from the asymmetry introduced by the different reproductive probabilities at individual level, no further asymmetry is assumed.

\subsubsection{The Genotype--Fitness Speciation Model}

Let us introduce evolution on genetic graphs as follows. In a community, individuals are labeled $i = 1,2,...,J_R$ $(J_P)$. Each individual $i$ can be described as belonging to a genetic group \cite{Lewontin&Kirk&Crow:1966}. In each genetic group, fitness of each individual $i$ within each species $k$ is given by the total number of individuals $j$ satisfying  $q^{ij}$  $>$ $q^{min}$, i. e., the total number of individuals each individual $i$ can mate with. Reproductive probability of individual $i$ within each species increases with the number of links or the number of genetically related mating partners (Fig. 2b). Thus the genetic level, using the genetic similarity among individuals, determines the speed and evolution of speciation rate and the genetic--species diversity. Each individual $i$ of species $k$ is chosen for reproduction with probability proportional to its fitness (see appendix):

\begin{equation}
 P_{i,k} = \frac{\sum_{j=1}^{N_{k}} H(q^{ij}-q^{min})}{S_R M_{k}}   
\end{equation}
where $H(\alpha)$ is
\begin{equation*}
H(\alpha) = \\
\begin{cases}
1 & \text{if $\alpha > 0$}\\
0 & \text{otherwise} 
\end{cases}
\end{equation*}
$N_{k}$, $S_{R}$ and $M_{k}$ are the abundance of species $k$, the total number of extant species in community $J_{R}$ and the total number of potential mating interactions within the species $k$, respectively. This genotype fitness model has the following same ingredient than the unified neutral model: (1) individuals have the same probability $1/J_R$ ($1/J_P$) to be chosen for death \cite{Tilman2:2004}, and (2) individuals equally connected within their own species or between species are equivalent in fitness, i.e, the identity to a given species does not confer per se fitness advantage, and the following additions: (1) fitness differences are then considered only within each species according to the number of genetically related mating partners; (2) there is density dependence across species, thus rare species have higher probabilities of reproduction, and (3) we select the most connected parents with higher probability which implies that the offspring can inherits their connectance, thus increasing its reproductive probability. Evolution selects for well connected individuals. In the same way, individuals are chosen for death and reproduction in the second community.

\subsubsection{The Phenotype--Fitness Speciation Model}

Let us now introduce evolution on ecological graphs as follows. In this last model, the phenotype class can be defined at the ecological level. Fitness of each individual $i$ within each population $k$ is associated with specific behavioral, morphological traits or phenotypic plasticity. Fitness is given by the total number of trophic links individual $i$ of population $k$ in one community has with $j$ individuals belonging to populations of the other community (Fig. 2c). In this phenotype fitness model the evolution of the connectivity at the individual level within each species determines the properties of the evolving system. The reproductive probability of individual $i$ increases with the number of trophic links but it is independent of its number of genetically related matings. Then, each individual $i$ of species $k$ is chosen for reproduction with probability proportional to its fitness (see appendix):

\begin{equation}
 P_{i,k} = \frac{\sum_{j=1}^{J_{P}} m_{ij}}{S_R M_{k}}   
\end{equation}
where the sum until $J_{P}$ means the total number of interactions of individual $i$ with all the individuals of community $J_{P}$. $m_{ij}$ means that there is an interaction between individual $i$ and $j$. $S_{R}$ and $M_{k}$ are the total number of extant species in community $J_{R}$, and the total number of mutualistic interactions among all the individuals of species $k$ with all the individuals in community $J_{P}$, respectively. This phenotype fitness model has the same two ingredients to the neutral and genotype model: (1) individuals have the same probability $1/J_R$ ($1/J_P$) to be chosen for death, and (2) individuals equally connected within their own species or between species are equivalent in fitness, i.e, the identity to a given species does not confer per se fitness advantage. The model has the following additions: (1) fitness differences are then considered only within each species according to the number of trophic links with individuals of the second community, (2) there is density  dependence in fitness across species, thus rare species have higher probabilities of reproduction, and (3) contrary to the genotype model, where the offspring inherits a number of potential matings from its parents, we assume that each offspring in this model starts with one trophic interaction. In the same way, individuals are chosen for death and reproduction in the second community.

\newpage

\bibliographystyle{plain}
\bibliography{MelianetalA}

\newpage
\section{Acknowledgments}
We thank Andrew P. Allen, Jordi Bascompte, Rick Condit, Scott Chamberlain, Jonathan Davies, Jennifer
Dunne, Rampal S. Etienne, Stanley Harpole,, Miguel A. Fortuna, Bill Langford,
Pablo Marquet, Neo Martinez, Mark Urban, C\'esar Vilas, Mark Vellend
and Tommaso Zillio for useful comments and ideas on the development of
the present study. CJM was supported by a Postdoctoral Fellowship at
the National Center for Ecological Analysis and Synthesis, a Center
funded by NSF (Grant \#DEB-0553768), the University of California,
Santa Barbara, and the State of California. CJM also acknowledges the support by Microsoft Research Ltd., Cambridge,
United Kingdom. DA acknowledges the support of the Netherlands
Organization for Scientific Research (NWO). DPV is a career researcher
with CONICET, and was also partly funded by FONCYT (PICT 20805).

\newpage
\section{Figure Legends}

$\bullet$ {Figure 1. The Higgs and Derrida model describes the stochastic evolution of a finite population of constant size. Individuals occupy the vertex of a graph. We start with a completely connected graph with two initial populations each with $J_R$ and $J_P$ individuals (Fig. 1a). A link between each pair of individuals denotes reproductive compatibility (i.e., $q^{ij}$ $>$ $q^{min}$). At each time step, an individual of each population is chosen for death (Fig 1b). Two individuals are chosen for reproduction. Individuals have the same probability $1/J_R$ ($1/J_P$) to be chosen for reproduction or for death (Fig 1c). The offspring of these two individuals replace the dead individual. The newborns in $J_{R}$ and $J_P$ can be the consequence of sexual reproduction without a mutualistic interaction (i.e., facultative mutualism given by $\omega$ $>$ 0) or a consequence of a mutualistic interaction with individual effectiveness $m$ between the first two chosen individuals for reproduction in population $J_{R}$ and $J_{P}$ (i.e., Fig. 1d, given by 1 - $\omega$).}

$\bullet$ {Figure 2a represents an example of the unified neutral model. In this example each community has 5 isolated groups with different number of individuals. The most abundant groups are interacting frequently, while the rare groups do not interact among them. This neutral model is the special case of an evolving multilevel graph with fitness of each individual according to the abundance of each population. Figure 2b and 2c represent a simple scenario for the genotype and phenotype models, respectively. In the genotype scenario an individual plant and a pollinator are linked according to the total number of genetically related mating partners each individual has in its population. For example, individuals represented with larger black circles in the $J_R$ and $J_P$ community have the highest number of mating links (3 in step 1). These individuals are selected as a parent of the offspring (not represented) and they are linked in step 2 (Fig. 2b). In the phenotype model an individual plant and pollinator are linked according to the total number of trophic interactions each individual has with individuals of the second community. For example, individuals with larger black circles in $J_R$ and $J_P$ have the highest number of trophic links (3 in step 1). These individuals are selected as the first parent of the offspring (not represented) and they are linked in step 2 (Fig. 2c).}

$\bullet$ {Figure 3 represents the distribution of Spearman's rank coefficient values with $J_R$ = $J_P$ = $10^{2}$, $\mu$ = $10^{-3}$, and $q^{min}$ equal to $0.9$ ($Q_{R}$ ($Q_{P}$) $\sim$  ${Q_{R}}^{*}$ (${Q_{P}}^{*}$) $\sim$ 0.7, see eq. 1). Fig. 3a,b,c represent the $J_R$ community under the neutral ($NUM$), the phenotype ($PF$) and the genotype ($GF$) scenarios, respectively. As expected, mating and trophic graphs are not correlated during the evolution of the system under the neutral and the phenotype fitness scenarios (i.e., the distribution of Spearman's p values are close to a uniform). We randomly sampled $10^{3}$ transient values from 10 replicates with the above mentioned parameter values. Fig. 3d,e represent the individual mating (Fig. 3d) and trophic (Fig. 3e) degree ranked from the most (left) to the least (right) connected individual after 9 randomly selected transients in one replicate from the genotype fitness model. Mating and trophic degree are correlated in the genotype fitness case. The distribution of the Spearman's values is highly skewed with approximately $80\%$ of p-values $<$ $0.01$.}

$\bullet$ {Figure 4a represents speciation events as a function of the number of generations for community $J_{R}$ (community $J_{P}$ not shown in the figure). Neutral, genotype and phenotype models are represented as circles, continuous and discontinuous lines, respectively. Each data point represents the average value after 100 replicates. We run each replicate for $10^{4}$ generations with $J_R$ = $J_P$ = $10^{3}$, $\mu$ = $10^{-4}$, and $q^{min}$ equal to $0.9$ ($Q_{R}$ ($Q_{P}$) $\sim$  ${Q_{R}}^{*}$ (${Q_{P}}^{*}$) $\sim$ 0.7, see eq. 1). Speciation events scale with the number of generations for all the three models ($r^2$ = 0.99, red lines) with the scaling exponent $\gamma$ = 0.97 (neutral), 1.03 (phenotype fitness) and 1.31 (genotype fitness). Fig. 4b represents the cumulative distribution of the number of generations to speciation for the neutral (circles), the genotype (continuous line) and the phenotype fitness models (discontinuous line). The distributions were generated using the mean of the sorted from the smallest to the largest number of generations to speciation after 100 replicates using the same parameter values than for Fig. 4a. On average, neutral evolution generates five and twice as many speciation events as evolving genetic and ecological networks with fitness differences, respectively.}

$\bullet$ {Figures 5a,b,c represent the evolution of the genetic ($1 - {Q_{R}}$), species diversity ($S_{e_{R}}$ following eq. 10), and species richness for the neutral (circles), the genotype (continuous line), and the phenotype fitness model (red line) with the number of generations. $J_R$ = $J_P$ = $10^{3}$, $\mu$ = $10^{-4}$, and $q^{min}$ equal to $0.9$ ($Q_{R}$ ($Q_{P}$) $\sim$  ${Q_{R}}^{*}$ (${Q_{P}}^{*}$) $\sim$ 0.7, see eq. 1 (results for $J_{P}$ not shown are qualitatively similar). On average, neutral evolution generates twice as many genetic (Fig. 5a) and species (Fig. 5b) diversity as evolving genetic and ecological networks with fitness differences, respectively. On the other hand, on average it generates five and twice as many number of coexisting species as evolving genetic and ecological networks with fitness differences, respectively (Fig. 5c). Results are the mean of $10^{4}$ values (i.e., one value per generation) after 100 replicates with the above mentioned parameter values.}

\newpage
\section{Figures}
\begin{figure}
\centering
\includegraphics[width=8cm]{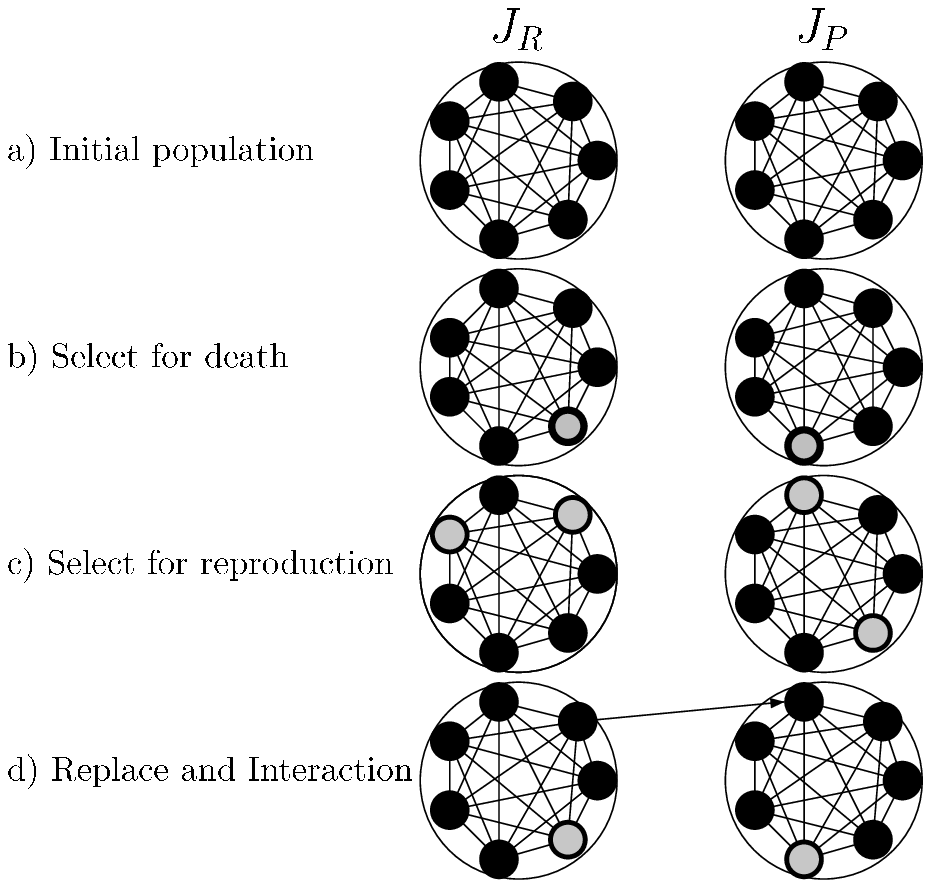}
\caption{}
\end{figure}
\begin{figure}
\centering
\includegraphics[width=14cm]{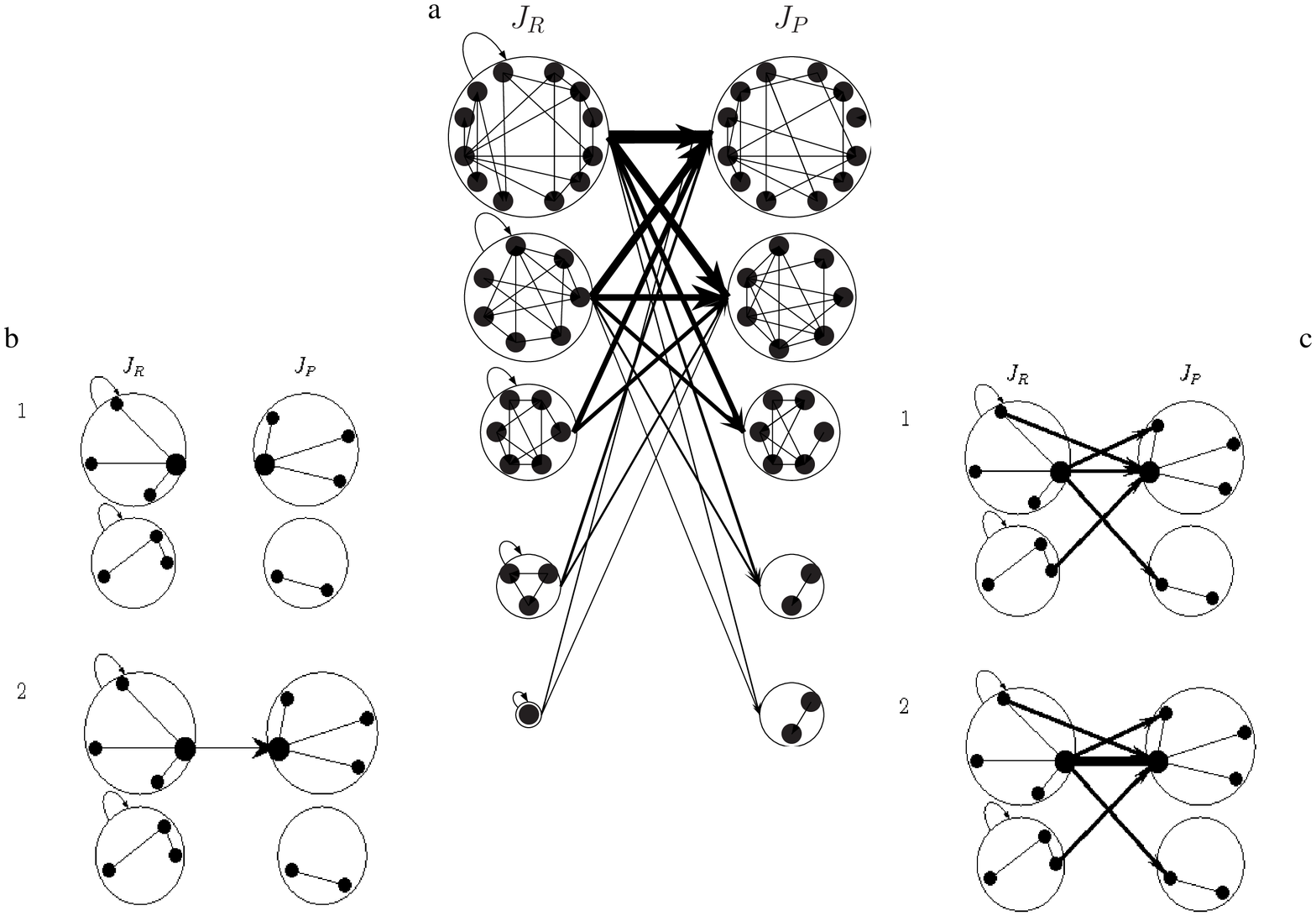}
\caption{}
\end{figure}
\begin{figure}
\centering
\includegraphics[width=18cm]{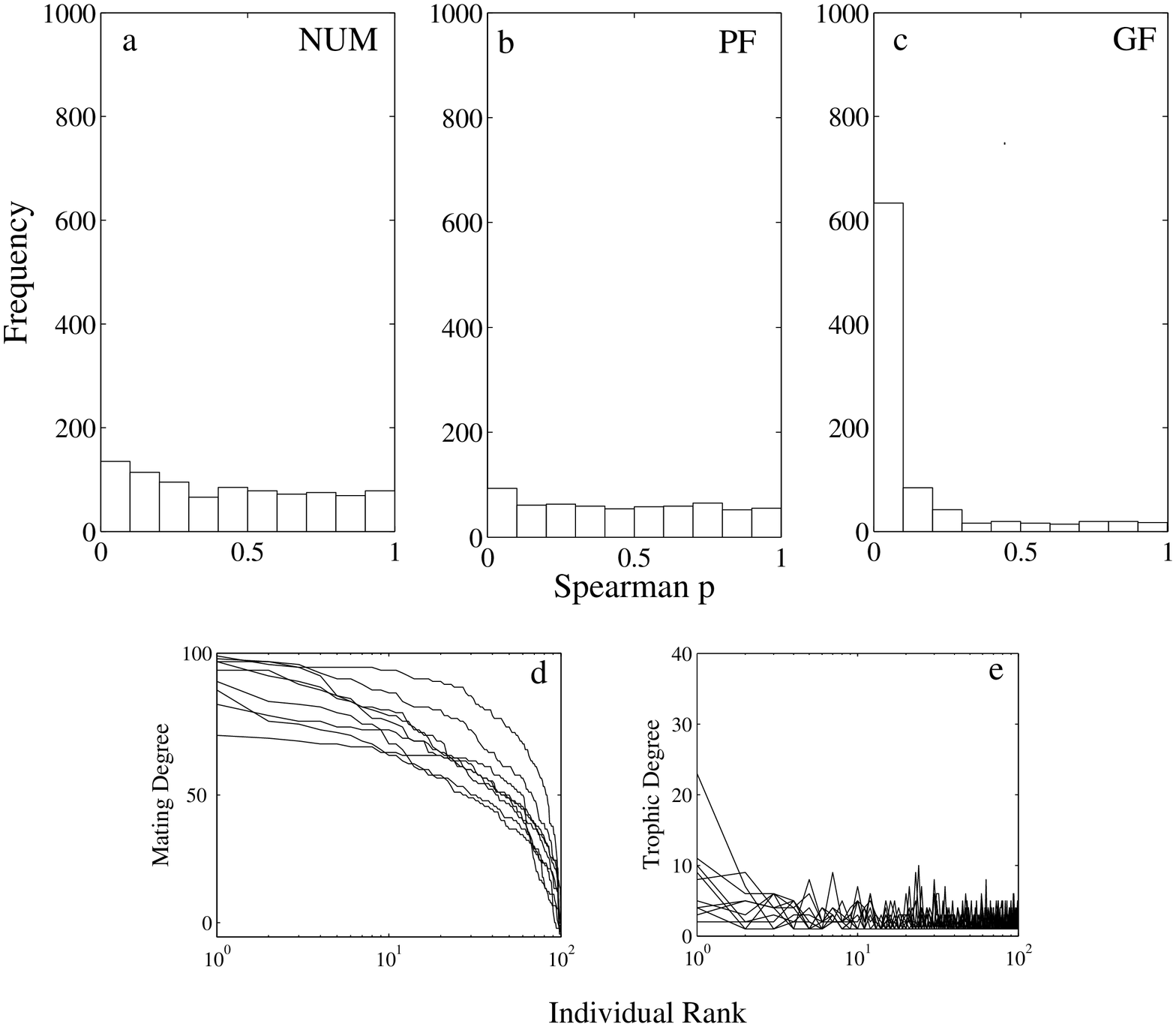}
\caption{}
\end{figure}
\begin{figure}
\centering
\includegraphics[width=16cm]{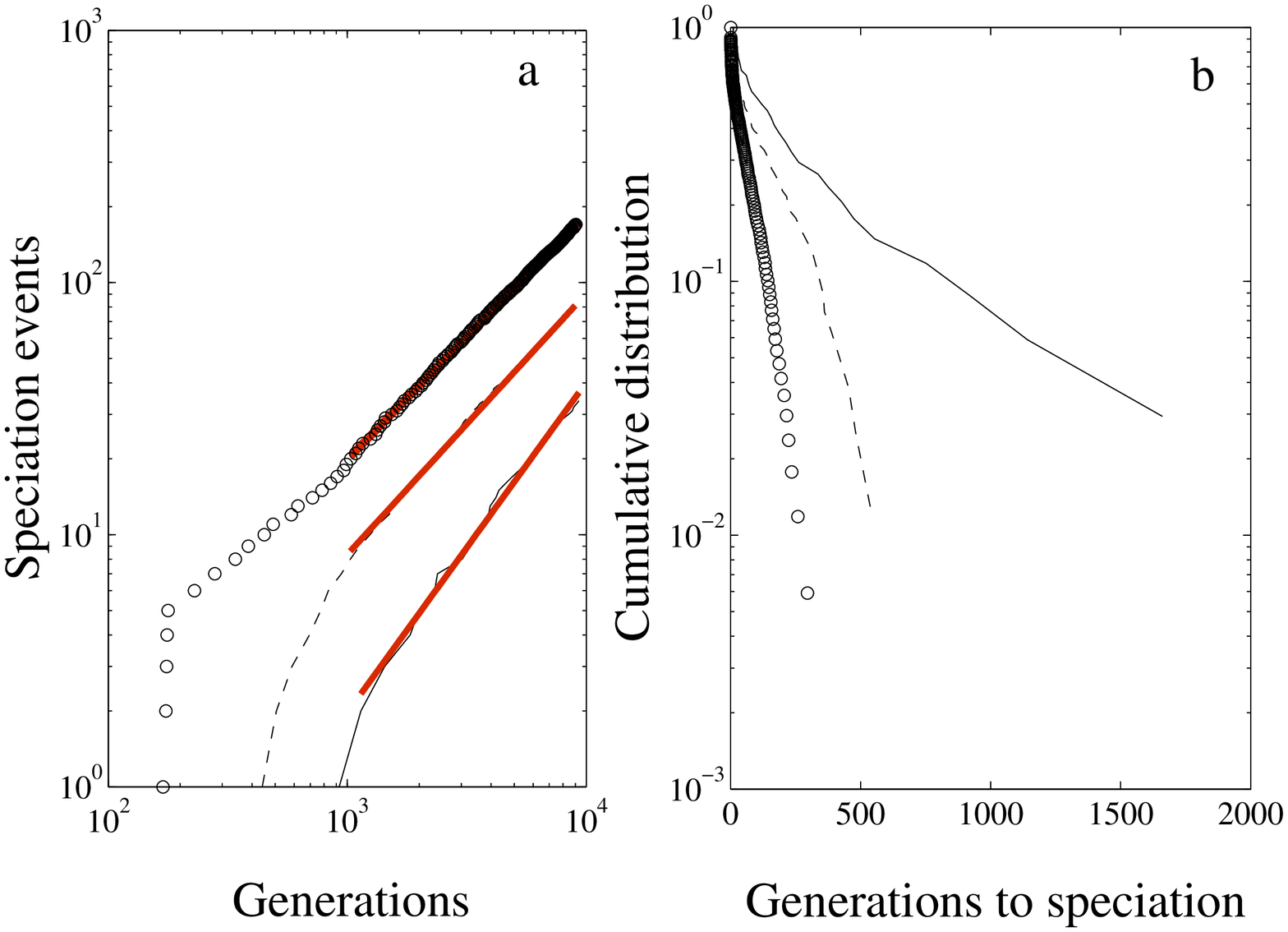}
\caption{}
\end{figure}
\begin{figure}
\centering
\hspace{-0.75 in}\includegraphics[width=18cm]{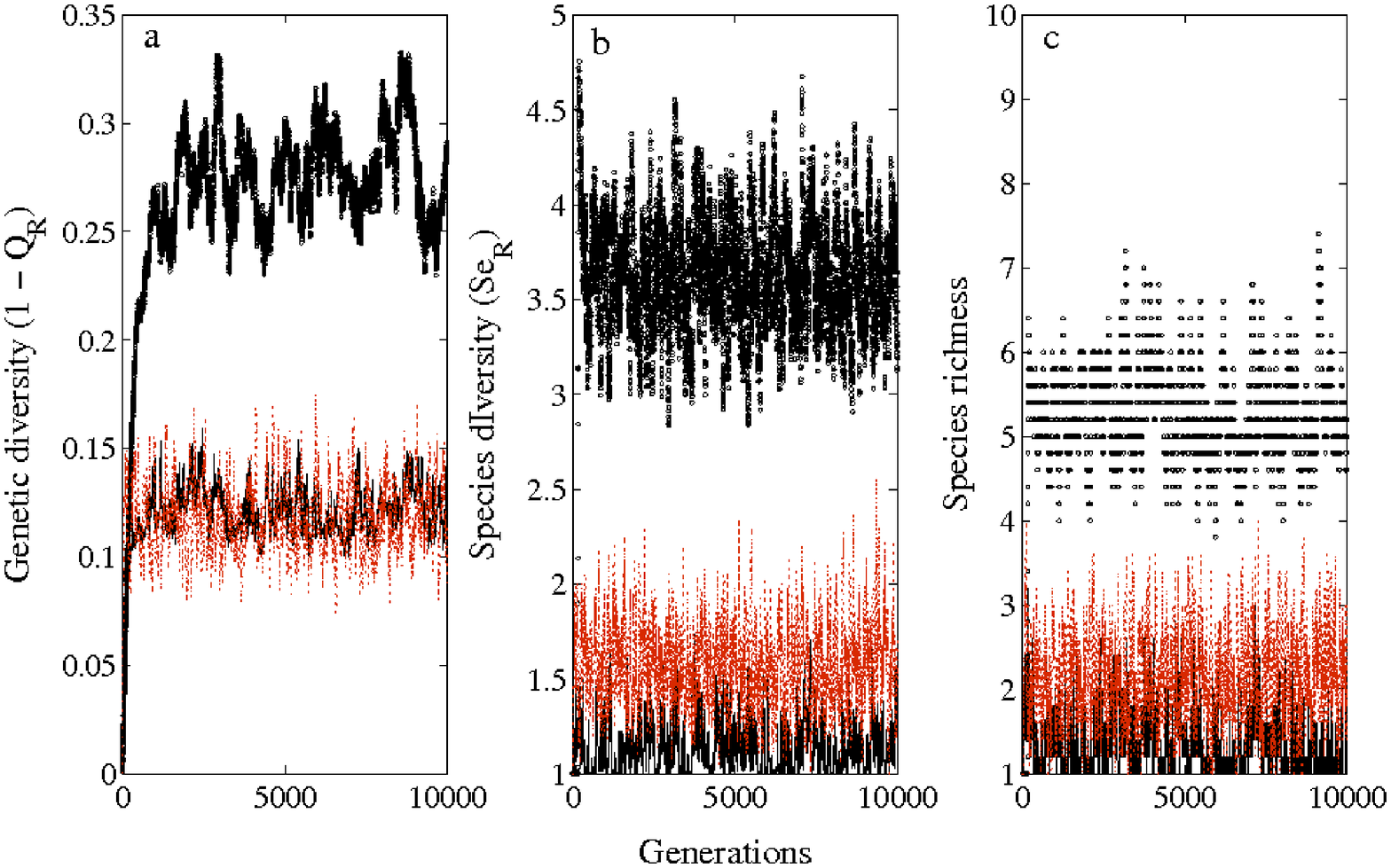}
\caption{}
\end{figure}
\end{document}


\maketitle
\baselineskip=8.75 mm

\linenumbers
\newpage

\appendix
\renewcommand{\theequation}{A-\arabic{equation}}
\setcounter{equation}{0}
\section{Appendix}

We first describe the model at molecular level starting by the process
of mutation, coancestry in overlapping generations and speciation. In
the end of the first part we describe in more detail the effective
number of alleles using the solution from the genetic similarity
matrix in a randomly mating population (section A.1). Second we
describe the effective number of species from the biodiversity theory
(section A.2). Third, we present a preliminary mathematical
description of our unified neutral model by using a master equation 
approach (section A.3). Fourth, we give a thorough description of our 
genotype (section A-4) and phenotype (section A-5) fitness speciation
model. Fifth, we provide further information about how we sampled the transients and the steady
state of our simulation models (section A.6). We also have included in section A.7 and A.8 the table legend for the tables A1, A2, A3 and A4. Table 1 clarifies the acronyms used in the main ms. Tables 2-4 clarify the different levels and mechanisms that can be considered in a general multilevel unified model. 

\subsection{Neutral Molecular Evolution Model: Mutation, Coancestry in Overlapping Generations and Speciation}

In the section ``Neutral Molecular Evolution Model'' of main text of
this paper we described briefly the three main components of the Higgs and Derrida
model: mutation ($\mu$), mating with random mixing of genes and
genetic drift. Here we attempt to describe in more detail those components. Let $\mu$ be the average rate of mutation of the alleles existing in a diploid population. Mutation rates are equal for forward and backward mutations and across loci. We consider one initial completely connected population with all sexual identical diploids $J$ individuals. 

Each individual $i$ is represented by a sequence of $N$ alleles each of which has two possible forms ($+1/-1$): (${S_{1}^{i},S_{2}^{i}...S_{N}^{i}}$), where $S_{u}^{i}$ is the $u^{th}$ unit in the sequence of individual $i$ \cite{Higgs&DerridaA:1992}. The initial genetic similarity values between each pair of individuals ($q^{ij}$) in the genetic similarity matrix $Q = [q^{ij}]$ are equal to 1. The genetic similarity between individual $i$ and individual $j$ is defined as

\begin{equation}
q^{ij}  = \frac{1}{N} \sum_{u=1}^{N} S_{u}^{i} S_{u}^{j}, \label{A1}
\end{equation}

At each time step, an individual is chosen for death. A second individual and its partner are chosen for reproduction. Individuals have the same reproductive probability ($1/J$) to be chosen for reproduction or for death. If $q^{ij}$ $>$ $q^{min}$, then the offspring of the two chosen parents replace (i.e., it occupies the empty site) the dead individual \cite{MayrA:1970,Higgs&DerridaA:1992,WuA:2001,Gavrilets2A:2004}. Otherwise the individuals do not mate (prezygotic $RI$) or their offspring is inviable or sterile (postzygotic $RI$) (i.e., we disregard the second individual and put it back). It is interesting to note here that this dynamics of speciation is derived from the underlying microevolutionary processes rather than postulated to follow a certain statistical distribution \cite{Hubbell2A:2001,Gavrilets2A:2004}. 

 Which is the expected sequence of each new offspring? Each new individual (k) has two parents $G_{1}(k)$ and $G_{2}(k)$. Each allele is inherited at random from one or other of the parent, thus ignoring linkage between neighboring alleles, but with a small probability of error determined by the mutation rate ($\mu$). Thus if $\mu$ $\rightarrow$ 0, n $\mu$ $\rightarrow$ $\lambda$, then as n $\rightarrow$ $\infty$,

\begin{equation}
(1 - \mu)^{n} \rightarrow e^{-\lambda}\hspace{0.5 in}
\end{equation}

and the probability to have in the unit of the sequence $S_{u}^{k}$ the same allele than one of its parents is
\begin{eqnarray}
P_{1}(S_{u}^{k} &=&S_{u}^{G_{1}(k)}) = \frac{1}{4}(1 + e^{-2\mu})\\
P_{2}(S_{u}^{k} &=&-S_{u}^{G_{1}(k)}) = \frac{1}{4}(1 - e^{-2\mu})\\
P_{3}(S_{u}^{k} &=&S_{u}^{G_{2}(k)}) = \frac{1}{4}(1 + e^{-2\mu})\\
P_{4}(S_{u}^{k} &=&-S_{u}^{G_{2}(k)}) = \frac{1}{4}(1 - e^{-2\mu}),\label{A2}
\end{eqnarray}
 
with the probabilities $P = (P_{1},P_{2},P_{3}, P_{4})$ satisfying $0$ $\leq$ $P_{i}$ $\leq$ $1$, $i = 1,2,..., \sum_{i}^{P_{4}} {P_{i}} = 1$. Given the value of the allele $S_{u}^{G_{1}(k)}$ of one of the parents, the expected value of that allele in the offspring is $E[S_{u}^{k}] =  e^{-2\mu} S_{u}^{G_{1}(k)}$. Similarly, given the similarity between each parent of the new individual $j$ and the individual $i$ already in the population we update the similarity matrix $q$ according to the following equation;
\begin{equation}
q^{ji}  = \frac{e^{-4\mu} }{2} (q^{G_{1}(j) i} + q^{G_{2}(j) i}).\label{A3}
\end{equation}
We know the evolution of the overlap matrix in the limit $N$ $\rightarrow$ $\infty$ in the infinite genome limit \cite{Higgs&DerridaA:1992}, because each pair of alleles contributing to $q^{ji}$ comes with equal probability from one of the two possible combinations of the parents of $j$ and individual $i$. By analyzing the similarity matrix $Q$ at any given time it is possible to assign new individuals to a species. The analysis of the similarity matrix $Q$ can be done by finding the isolated subgroups of individuals. In each time step we first check the $k$ individuals that can mate with the newborn $j$ (i.e., $q^{jk}$ $>$ $q^{min}$). Second we check all $k$ individuals that can mate with $j$ but can not mate with the rest of $i$ individuals in the population ($q^{ik}$ $\leq$ $q^{min}$ for all $i$). If this two criteria are satisfy, then we have an isolated group.

The probability that two randomly selected individuals from the genetic similarity matrix $Q$ have a common ancestor at time $t$ is (see \cite{MoranA:1962,FelsensteinA:1971,JohnsonA:1977,Choy&WeirA:1978} for a detailed discussion about the similarities and differences with respect to monoecious and dioecious diploid populations with overlapping generations) 

\begin{eqnarray}
Q_{t} = e^{-4\mu} \Biggl[\frac{1}{2 J} + \Biggl(1-\frac{1}{2 J}\Biggl) Q_{t-1} \Biggl]\\ \nonumber
\end{eqnarray}

If $\mu$ $<<$ $1$, then the genetic similarity matrix $Q$ has a solution \cite{Higgs&DerridaA:1992}:
\begin{eqnarray}
{Q}^{*} = \frac{1}{\theta_{m} + 1}\\ \nonumber
\end{eqnarray}
where $\theta_{m}$ = $4 J \mu$. This solution is identical to the inbreeding coefficient ($F$) in population genetics with non-overlapping generations \cite{WrightA:1931,Kimura&CrowA:1964} meaning the probability that an individual will be homozygous. It is interesting to note that ${Q}^{*}$ and $F$ have the same value despite that ${Q}^{*}$ is an average property of all loci in the sequence whereas $F$ is defined for the infinite allele model in a single locus \cite{Higgs&DerridaA:1992}. The probability that two individuals do not have a common ancestor is given by
\begin{eqnarray}
1 - {Q}^{*} =  \frac{\theta_{m}}{\theta_{m} + 1}\\ \nonumber
\end{eqnarray}
where $1 - {Q}^{*}$ is a measure of heterozygosity or genetic diversity.  As we commented in the main part of the ms. if $q^{min}$ $>$ ${Q}^{*}$, then the initial population $J$ is greatly perturbed by the cutoff, the connectivity within the population decreases and the genetic similarity matrix (${Q}$) can never reach its equilibrium state. Initial population splits and speciation happens with the species fluctuating in the system according to demographic stochasticity \cite{Higgs&DerridaA:1992}.

\subsection{Neutral Theory of Biodiversity: The Effective Number of Species}

Neutral theory of biodiversity discusses species instead of alleles. In this case, the probability that two individuals in population $J$ chosen at random will be of the same species is

\begin{eqnarray}
f_{t} = (1-\nu)^2 \Biggl[\frac{1}{J} + \Biggl(1-\frac{1}{J}\Biggl) f_{t-1} \Biggl]\\ \nonumber
\end{eqnarray}

At equilibrium, assuming $\nu$ is small, we have

\begin{eqnarray}
{f}^{*} = \frac{1}{\theta_{b} + 1}\\ \nonumber
\end{eqnarray}
where  $\theta_{b}$ = $J \nu$. Then, the {\em effective number of species} ($S_{e}$) in the $J$ community is \cite{NeeA:2005}

\begin{eqnarray}
S_{e} = \theta_{b} + 1\\ \nonumber
\end{eqnarray}
which represents a measure of species diversity and is identical to the Simpson's species diversity index \cite{SimpsonA:1949,NeeA:2005}. In this scenario ecological drift dominates and speciation is an individual level process that leads to a proportional relationship between the speciation rate of each species in a community and its abundance \cite{Hubbell2A:2001}. How can we link neutral molecular and community ecology theories using evolving multilevel networks? In this section we describe in more detail the neutral unified model.

\subsection{Neutral Unified Model: A Master Equation Approach with Explicit Speciation in Two Interacting Communities.}

Our main goal here is to describe in further detail our basic neutral 
model. In order to do it, we provide a preliminary mathematical
description of the evolutionary and ecological processes controlling 
community dynamics. An important point is that two initial populations 
give rise to two interacting communities, i. e., through a combined
effect of speciation, death, and reproduction an ecological network 
connecting two mutualistic communities emerges. Model dynamics is controlled by three input parameters: the mutation rate ($\mu_{R}$ = $\mu_{P}$ = $\mu$), the minimum genetic similarity value ($q^{min}_{R}$ = $q^{min}_{P}$ = $q^{min}$) at genetic level, ultimately controlling the speciation process, and the individual mutualistic effectiveness ($m$) assuming a fully symmetric case for all the individuals interactions at the ecological level. 

Our simulations consider a zero-sum birth and death stochastic
individual based model in sexual diploids populations with overlapping
generations and age independent birth and death rates in the context of neutral mutations and large genome size
per individual (effectively infinite gene sequences) 
\cite{Higgs&DerridaA:1992,Gavrilets2A:2004}. 
Individual interactions are introduced using a single and
large-homogeneous patch (or metacommunity) in which there is a complete
mixing and all individuals have the same chance of potentially
interacting with each other \cite{Hubbell2A:2001,McKane1A:2004}. Our
model produces two mutualistically interacting communities 
---the resource $R$ or plant community and the consumer or pollinator
$P$ community--- but it can be easily extended to any kind of 
ecological interactions and to a larger number of interacting communities. 

The two communities assume hermaphroditic plant and
dioecious pollinator individuals. They are labeled $R={1,\dots,J_R}$,
and $P={1,\dots,J_P}$,  where $J_R$ and $J_P$ are the total number of
individual plants and pollinators, respectively. Site size inside
the patch for plants (pollinators) is defined so that each one
contains one $R$ ($P$) plant (pollinator) individual. Thus $J_{m}$ =
$J_R$ + $J_P$ is the total number of effective individuals in the
metacommunity which implies that all individuals are considered in the
reproductive age. These numbers are kept constant by assuming 
zero-sum dynamics. 

An important remark is that our simulations are run much faster 
under the zero-sum rule, but community dynamics is easier to describe
by dropping this assumption. Therefore, in what follows, for clarity
and simplicity, we assume that death and non-mutualistic reproduction 
can take place independently in either community. Only mutualistic 
interactions involve simultaneously individuals from the two communities. 
Zero-sum models are equivalent to their non zero-sum counterparts 
at stationarity \cite{EtiennetalA:2008}. Although we
have run all our simulations under zero-sum dynamics (see our code 
in section A.9), we are quite confident that our main results are
robust and do not rely on the specific implementation of the zero-sum 
rule. 

Consider two randomly mating populations of size $J_R$ and $J_P$ where
each individual has an infinitely large genome sequence subject to
random neutral mutations. The initial genetic similarity values
between each pair of individuals ($q^{ij}_{R}$ and $q^{ij}_{P}$) in
the matrices $Q_{R} = [q^{ij}_{R}]$ and  $Q_{P} = [q^{ij}_{P}]$ are
equal to 1 and mutation rates $\mu_{R}$ and $\mu_{P}$ are equal among
all individuals. Individuals have the same probability $1/J_R$
($1/J_P$) to be chosen for reproduction or for death. An individual of
each population is chosen for death and two individuals of each
population are chosen for reproduction. The offspring from the two
selected individuals for reproduction of the same population replace
the dying individual.  Parameter $\omega$ is defined as the
probability of facultative mutualistic interaction, i. e., the
probability of having sexual reproduction in plants without
pollinators and sexual reproduction in pollinators without plants. 
To be more precise, any $k$-species within the plant ($R$) and
pollinator ($P$) communities will undergo the following processes:

\begin{enumerate}
\item Death of an individual in the plant $R$ community:
\begin{equation}
\centerline{{\large $R^k \xrightarrow{d_{N^k_R}} \emptyset^R$}} \hspace{-1.8 in} \label{A4}
\end{equation}
\item Death of an individual in the pollinator $P$ community:
\begin{equation} 
\centerline{{\large $P^k \xrightarrow{d_{N^k_P}} \emptyset^P$}} \hspace{-1.8 in} \label{A5}
\end{equation}
\item Non-mutualistic reproduction of an individual in the $R$
  community:
\begin{equation}
\centerline{{\large $\emptyset^R \xrightarrow{b_{N^k_R} M^k_R (1 - P^k_R)} R^k$}} \label{A6} \hspace{-1.5 in} 
\end{equation} 
\item Non-mutualistic reproduction of an individual in the $P$
  community:
\begin{equation}
\centerline{{\large $\emptyset^P \xrightarrow{b_{N^k_P} M^k_P (1 - P^k_P)} P^k$}} \label{A7} \hspace{-1.5 in}
\end{equation}
\item Arising of a new species $j$ as a result of a
  non-mutualistic reproduction in the $R$ community: 
\begin{equation}
\centerline{{\large $\emptyset^R \xrightarrow{b_{N^k_R} M^k_R P^k_R} R^j$}} \label{A8} \hspace{-1.6 in}
\end{equation}
\item Arising of a new species $j$ as a result of a
  non-mutualistic reproduction in the $P$ community:
\begin{equation}
\centerline{{\large $\emptyset^P \xrightarrow{b_{M^k_P} M^k_P P^k_P} P^j$}} \label{A9} \hspace{-1.6 in}
\end{equation}
\end{enumerate}

Furthermore, in case of mutualism is obligate, reproduction and ecological
interaction are considered to be coupled events involving a pair of
individuals from each community. The strength of the mutualistic
interaction is controlled by $1 - \omega$, which is defined as an
individual mutualistic effectiveness, $m_{N^k_{R},N^k_{P}}$. Any time
that a mutualistic event occurs, an interacting link connecting the 
individuals involved in the interaction appears. It is in this way
that a dynamical ecological network connecting the two communities 
emerges. Notice that this reproduction-interaction coupled event can
result in four different outputs: 

\begin{eqnarray}
\centerline{ {\large \xymatrix @C=18pc {\emptyset^{R} \ar @/_0.25pc/[r]^{m_{N^{k}_{R},N^k_{P}} M^{k}_{R} (1 - P^{k}_{R}) M^{k}_{P} (1 - P^{k}_{P})} &R^{k}}}} \label{A10}
\\
\centerline{  {\large \xymatrix @C=18pc {\emptyset^{P} \ar @/^1.75pc/[r] &P^{k}}}} \nonumber
\end{eqnarray}

\begin{eqnarray}
\centerline{ {\large \xymatrix @C=18pc {\emptyset^{R} \ar @/_0.25pc/[r]^{m_{N^{k}_{R},N^{k}_{P}} M^{k}_{R} (1 - P^{k}_{R}) M^{k}_{P} P^{k}_{P}} &R^{k}}}} \label{A11}
\\
\centerline{  {\large \xymatrix @C=18pc {\emptyset^{P} \ar @/^1.75pc/[r] &P^{j}}}} \nonumber
\end{eqnarray}

\begin{eqnarray}
\centerline{ {\large \xymatrix @C=18pc {\emptyset^{R} \ar @/_0.25pc/[r]^{m_{N^{k}_{R},N^{k}_{P}} M^{k}_{R} P^{k}_{R} M^{k}_{P} (1 - P^{k}_{P})} &R^{j}}}} \label{A12}
\\
\centerline{  {\large \xymatrix @C=18pc {\emptyset^{P} \ar @/^1.75pc/[r] &P^{k}}}} \nonumber
\end{eqnarray}

\begin{eqnarray}
\centerline{ {\large \xymatrix @C=18pc {\emptyset^{R} \ar @/_0.25pc/[r]^{m_{N^{k}_{R},N^{k}_{P}} M^{k}_{R} P^{k}_{R} M^{k}_{P} P^{k}_{P}} &R^{j}}}} \label{A13}
\\
\centerline{  {\large \xymatrix @C=18pc {\emptyset^{P} \ar @/^1.75pc/[r] &P^{j}}}} \nonumber
\end{eqnarray}
\\
where $\emptyset^{R}$ ($\emptyset^{P}$), represent empty sites, 
and $d_{N^{k}_{R}}$ ($d_{N^{k}_{P}}$) and $b_{N^{k}_{R}}$ ($b_{N^{k}_{P}}$)  death and birth per
capita rates of plant and pollinator species $k$, respectively. $M^{k}_{R}$ and $M^{k}_{P}$ are the probabilities to pick up
randomly two individuals that can actually mate 
[($i^{'}$,$j^{'}$) with $q^{i^{'}j^{'}}$ $>$ $q^{min}$] among all available pairs 
in hermaphroditic plant species $N^{k}_{R}$ and pollinator species $N^{k}_{P}$, respectively:
\begin{equation}
M^{k}_{R} = P\{i^{'},j^{'} | q^{i^{'}j^{'}} > q^{min}\} = \frac{2}{N^{k}_{R}(N^{k}_{R} + 1)} \sum_{i^{'} = 1}^{N^{k}_{R}} \sum_{j^{'} = i^{'}}^{N^{k}_{R}} H(q^{i^{'}j^{'}} - q^{min}),
\end{equation}

\begin{equation}
M^{k}_{P} = P\{i^{'},j^{'} | q^{i^{'}j^{'}} > q^{min}\} = \frac{2}{N^{k}_{P}(N^{k}_{P} - 1)} \sum_{i^{'} = 1}^{N^{k}_{P}} \sum_{j^{'} = i^{'} + 1}^{N^{k}_{P}} H(q^{i^{'}j^{'}} - q^{min}),
\end{equation}

where $H(\alpha)$ 
\begin{equation*}
H(\alpha) = \\
\begin{cases}
1 & \text{if $\alpha > 0$}\\
0 & \text{otherwise} 
\end{cases}
\end{equation*}
\vspace{0.2 in}

and $\nu^{k}_{R}$ and $\nu^{k}_{P}$ are the probabilities of mutation-induced
speciation for species $k$ in the plant and pollinator populations, respectively:
\begin{equation}
\nu^{k}_{R} = \frac{2}{N^{k}_{R}(N^{k}_{R} + 1)} \sum_{k^{'}_{1} = 1}^{N^{k}_{R}} \sum_{k^{'}_{2} = k^{'}_{1}}^{N^{k}_{R}} P^{k^{'}_{1},k^{'}_{2}}_{R},
\end{equation}

\begin{equation}
\nu^{k}_{P} = \frac{2}{N^{k}_{P}(N^{k}_{P} - 1)} \sum_{k^{'}_{1} = 1}^{N^{k}_{P}} \sum_{k^{'}_{2} = k^{'}_{1} + 1}^{N^{k}_{P}} P^{k^{'}_{1},k^{'}_{2}}_{P},
\end{equation}
\vspace{0.2 in}

where $P^{k^{'}_{1},k^{'}_{2}}_{R}$ and $P^{k^{'}_{1},k^{'}_{2}}_{P}$ are defined as the
probabilities of producing an individual offspring belonging to a new
species, after two parent individuals ($k^{'}_{1},k^{'}_{2}$) have been randomly
chosen from plant and pollinator $k$ species, respectively. This new
individual belongs to a new species provided it is sexually
incompatible with any other individual in the respective community:

\begin{equation}
{\Large P^{k^{'}_{1},k^{'}_{2}}_{R} = F\left[ \frac{\sum_{\substack{i = 1\\ i \neq j}}^{J_{R}} H(q_{t} - (q^{k^{'}_{1} i} + q^{k^{'}_{2} i}))}{J_{R} - 1} \right]}
\end{equation}
\begin{equation}
{\Large P^{k^{'}_{1},k^{'}_{2}}_{P} = F\left[ \frac{\sum_{\substack{i = 1\\ i \neq j}}^{J_{P}} H(q_{t} - (q^{k^{'}_{1} i} + q^{k^{'}_{2} i}))}{J_{P} - 1} \right]}
\end{equation}
\vspace{0.2 in}

where $q_{t}$ = $\frac{2 q^{min}}{e^{-4 \mu}}$, $F[x] = 1$ if $x=1$ and zero otherwise $H(\alpha)$ is 
\begin{equation*}
H(\alpha) = \\
\begin{cases}
1 & \text{if $\alpha > 0$}\\
0 & \text{otherwise} 
\end{cases}
\end{equation*}

Expressions above characterize speciation. $P^{k^{'}_{1},k^{'}_{2}}_{R}$ will be 
either 1 when speciation occurs and zero otherwise. When the offspring 
is a new individual $i$ that cannot mate with any individual in the
community (with $i$ $\neq$ $j$), we have an incipient species of size
1. This a mutation-induced speciation event. However, note 
that there is also an alternative speciation mode. Death events may 
induce the formation of new species of larger sizes. When there is
only one individual connecting two mating subnetworks within the same 
species and this ``connector''individual happens to die, the ancestral 
species splits into two daughter species, which implies that an
additional new species arises. This is a fission-induced speciation
event. In order to fully characterize this second speciation mode, 
we would require to count the number of ``connector'' individuals
per species and the distribution of subnetwork sizes those individuals
are connecting at any point in time. 

In order to write down an equation describing the temporal dynamics
of the two interacting communities, we need to characterize the state 
of the system and all possible transitions between states at a given
time. Transition rates naturally follow from the set of events we have 
considered above. The state of the system is defined by the community 
abundance vectors for the plant 
$\overrightarrow{N_{R}}$ = $[N^{R}_{1}, N^{R}_{2},N^{R}_{3},...N^{R}_{S_R}]$ 
and pollinator 
$\overrightarrow{N_{P}}$ = $[N^{P}_{1}, N^{P}_{2},N^{P}_{3},...N^{P}_{S_P}]$
community, and by the two genetic similarity matrices of dimensions
$J_{R} \times J_{R}$ and $J_{P} \times J_{P}$ corresponding to the two 
communities which control the evolving mating networks. These
matrices ultimately control reproduction and speciation rates through
the probabilities $M^{k}_{*}$ and  $P^{k}_{*}$, where $*$ stands for $R$ and $P$. 
Notice that in the limit of large genome sizes there is no
stochasticity in the similarity matrices. They are updated following
the rule given by Eq. (\ref{A3}) after any single reproduction, 
extinction or speciation event.

We study a fully symmetric case ($m_{N^{k}_{R},N^{k}_{P}}$~=~$m_{N^{k}_{P},N^{k}_{R}}$) 
between plant and pollinator individuals, within species, and across species within 
the plant and pollinator community. In fact, for simplicity, we
also assume $m_{N^{k}_{R},N^{k^{'}}_{P}}$~=~$m_{N^{k}_{P},N^{k^{'}}_{R}}$~=~$m$~=~1,
i. e., the obligate mutualistic scenario, and percapita
birth and death rates are assumed to take the same value across 
species and within and across the two communities 
($d^{k}_{R}$ = $d^{k^{'}}_{P}$ and $b^k_{R}$ = $b^{k^{'}}_{P}$). 
This will further simplify transition rates by scaling time
according to the birth-death temporal rate (a single time step is the 
time required for one death and one birth per community to occur).  

We define $\overrightarrow{e^{*}_{k}}$ = $[e_{1}^{k}, \dots, e_{S_*}^{k}]$, where $*$ stands for $R$ and $P$ and $e_{i}^{k}$ is 1 if $k=i$
and 0 otherwise. The events considered in (\ref{A4})-(\ref{A13}) allow
to define the following transition rates. Note these three remarks 
in order. First, the system can only loose one individual from a given species 
either through a death in the plant or pollinator community and this transition
probability rate should increase linearly with the abundance of that species. 
Second, the encounter of two individuals for reproduction 
is a quadratic process that involve the abundance of that species
squared. Third, the speciation process increases the dimension of the
abundance community vector with the addition of a new component
corresponding to the new incipient species. With this in mind, we can 
readily write:

\begin{enumerate}

\item Death:
\begin{multline}
\hspace{-0.5 in} T \left[\overrightarrow{N_{R}} - \overrightarrow{e^{R}_{k}}, \overrightarrow{N_{P}} \hspace{0.1 in} | \hspace{0.1 in} \overrightarrow{N_{R}}, \overrightarrow{N_{P}}\right] = \left(\frac{N^{k}_{R}}{J_{R}}\right) \label{A14}
\end{multline}
\begin{multline}
\hspace{-0.5 in} T \left[\overrightarrow{N_{R}}, \overrightarrow{N_{P}} - \overrightarrow{e^{P}_{k}} \hspace{0.1 in} | \hspace{0.1 in} \overrightarrow{N_{R}}, \overrightarrow{N_{P}}\right] = \left(\frac{N^{k}_{P}}{J_{P}}\right) \label{A15}
\end{multline}

\item Non-mutualistic reproduction:
\begin{multline}
\hspace{-0.5 in} T \left[\overrightarrow{N_{R}} + \overrightarrow{e^{R}_{k}}, \overrightarrow{N_{P}} \hspace{0.1 in}
  | \hspace{0.1 in} \overrightarrow{N_{R}},\overrightarrow{N_{P}}\right] = \omega M^{k}_{R}\frac{(N^{k})^{2}}{J_{R} J_{R}} \left(1 - P^{k}_{R}\right)
\label{A16}
\end{multline}
\begin{multline}
\hspace{-0.5 in} T \left[\overrightarrow{N_{R}} , \overrightarrow{N_{P}} + \overrightarrow{e^{P}_{k}} \hspace{0.1 in} 
  | \hspace{0.1 in} \overrightarrow{N_{R}},\overrightarrow{N_{P}}\right] = \omega M^{k}_{P}\frac{(N^{k})^{2}}{J_{P} J_{P}}\left(1 - P^{k}_{P}\right)
\label{A17}
\end{multline}

\item Speciation associated to non-mutualistic reproduction:
\begin{multline}
\hspace{-0.5 in} T \left[\overrightarrow{N_{R}} + \overrightarrow{e^{R}_{S_R +1}}, \overrightarrow{N_{P}} \hspace{0.1 in}
  | \hspace{0.1 in} \overrightarrow{N_{R}},\overrightarrow{N_{P}}\right] = \omega M^{k}_{R}\frac{(N^{k})^{2}}{J_{R} J_{R}} \left(P^{k}_{R}\right)
\label{A18}
\end{multline}
\begin{multline}
\hspace{-0.5 in} T \left[\overrightarrow{N_{R}} , \overrightarrow{N_{P}} + \overrightarrow{e^{P}_{S_P+1}} \hspace{0.1 in} 
  | \hspace{0.1 in} \overrightarrow{N_{R}},\overrightarrow{N_{P}}\right] = \omega M^{k}_{P}\frac{(N^{k})^{2}}{J_{P} J_{P}}\left(P^{k}_{P}\right)
\label{A19}
\end{multline}

\item Mutualistic reproduction:
\begin{multline}
\hspace{-0.5 in} T \left[\overrightarrow{N_{R}} + \overrightarrow{e^{R}_{k}}, \overrightarrow{N_{P}} + \overrightarrow{e^{P}_{k^{'}}} \hspace{0.1 in} | \hspace{0.1 in} \overrightarrow{N_{R}},\overrightarrow{N_{P}}\right] = (1-\omega) M^{k}_{R}\frac{(N^{k})^{2}}{J_{R} J_{R}} \left(1 - P^{k}_{R}\right) M^{k^{'}}_{P}\frac{(N^{k^{'}})^{2}}{J_{P} J_{P}}\left(1 - P^{k^{'}}_{P}\right) 
\label{A20}
\end{multline}

\item Speciation associated to mutualistic reproduction furnishes the three remaining transitions (either or both species involve in the interaction undergo speciation):
\begin{multline}
\hspace{-0.5 in} T \left[\overrightarrow{N_{R}} + \overrightarrow{e^{R}_{S_R +1}}, \overrightarrow{N_{P}} + \overrightarrow{e^{P}_{k'}} \hspace{0.1 in} | \hspace{0.1 in} \overrightarrow{N_{R}},\overrightarrow{N_{P}}\right] = (1-\omega) M^{k}_{R}\frac{(N^{k})^{2}}{J_{R} J_{R}} \left(P^{k}_{R}\right) M^{k^{'}}_{P}\frac{(N^{k^{'}})^{2}}{J_{P} J_{P}}\left(1 - P^{k^{'}}_{P}\right) 
\label{A21}
\end{multline}
\begin{multline}
\hspace{-0.5 in} T \left[\overrightarrow{N_{R}} + \overrightarrow{e^{R}_{k}}, \overrightarrow{N_{P}} + \overrightarrow{e^{P}_{S_P+1}} \hspace{0.1 in} | \hspace{0.1 in} \overrightarrow{N_{R}},\overrightarrow{N_{P}}\right] = (1-\omega) M^{k}_{R}\frac{(N^{k})^{2}}{J_{R} J_{R}} \left(1 - P^{k}_{R}\right) M^{k^{'}}_{P}\frac{(N^{k^{'}})^{2}}{J_{P} J_{P}}\left(P^{k^{'}}_{P}\right) 
\label{A22}
\end{multline}
\begin{multline}
\hspace{-0.5 in} T \left[\overrightarrow{N_{R}} + \overrightarrow{e^{R}_{S_R+1}}, \overrightarrow{N_{P}} + \overrightarrow{e^{P}_{S_p+1}} \hspace{0.1 in} | \hspace{0.1 in} \overrightarrow{N_{R}},\overrightarrow{N_{P}}\right] = (1-\omega) M^{k}_{R}\frac{(N^{k})^{2}}{J_{R} J_{R}} \left(P^{k}_{R}\right) M^{k^{'}}_{P}\frac{(N^{k^{'}})^{2}}{J_{P} J_{P}}\left(P^{k^{'}}_{P}\right) 
\label{A23}
\end{multline}  

\end{enumerate}

As we have done in all our simulations, if we consider only the
obligate mutualistic scenario, $\omega = 0$, or mutualistic 
effectiveness, $m_{N^{k}_{R},N^{k}_{P}}$ = $m_{N^{k}_{P},N^{k}_{R}}$ is equal
to 1, stochastic transition rates simplify and system is described
by death rates (\ref{A14})-(\ref{A15}), mutualistic pure 
reproduction (\ref{A20}), and the speciation rates associated to 
obligate mutualistic reproduction, (\ref{A21})-(\ref{A23}).

It is important to remark that the stochastic events we have 
considered are neglecting the fission speciation mode. However, 
all results presented in the main ms. are based on a zero-sum 
code which does take into account both
mechanisms of speciation. Therefore, our preliminary mathematical 
description above provides only an approximation to the actual 
dynamics of the system. Within this limitation, we have provided a 
set of transition probabilities that allow to build exact stochastic 
simulations \cite{Gillespie:1976}, and a master
equation that provides a general description of the time evolution of
two interacting communities with explicit speciation and mutualistic interactions.
These tools promise to expand our ability for quantitative analysis. 
However, more work is needed to fully characterize the contribution to the
formation of new species through the fission-mode speciation 
mechanism.  

\subsection{Genotype Fitness Speciation Model}

Do genetic or ecological mechanisms drive the speed of speciation,
genetic--species diversity and coexistence? 
We here describe in detail the individual fitness according to its
reproductive probabilities. Each individual $i$ of species $k$ is
chosen for reproduction with probability proportional to its fitness,

\begin{equation}
 P_{i,k} = {\cal{N}} F_{i,k}
\end{equation}
where individual fitness is defined as:
\begin{equation}
  F_{i,k} = \frac{\sum_{j=1}^{N_{k}} H(q^{ij}-q^{min})}{M_{k}}   
\end{equation}
Thus, we write:
\begin{equation}
  P_{i,k} = {\cal{N}} \frac{\sum_{j=1}^{N_{k}} H(q^{ij}-q^{min})}{M_{k}}   
\end{equation}
where $\cal{N}$ is a normalization factor, $N_{k}$ is the abundance of
species $k$, and $M_{k}$ is the total number of potential mating
interactions within the species $k$, which, in turn, can be written as:
\begin{equation}
  M_{k} = \sum_{i = 1}^{N_{k}} \sum_{j = 1}^{N_{k}} H(q^{ij} - q^{min})   
\end{equation}
and $H(\alpha)$ 
\begin{equation*}
H(\alpha) = \\
\begin{cases}
1 & \text{if $\alpha > 0$}\\
0 & \text{otherwise} 
\end{cases}
\end{equation*}

We now calculate the normalization factor by using the normalization
requirement, i.e., by summing $P_{i,k}$ across all individuals 
and species 1 must be obtained:
\begin{equation}
  {\cal{N}} \sum_{i=1}^{S_{R}} \sum_{j=1}^{N_{k}} F_{i,k} = 1   
\end{equation}
and, then:
\begin{equation}
  {\cal{N}} = \frac{1}{\sum_{i=1}^{S_{R}} \sum_{j=1}^{N_{k}} F_{i,k}}
\end{equation}
Therefore, the probability of birth for each $i$ individual is:
\begin{equation}
  P_{i,k} = \frac{F_{i,k}}{\sum_{i=1}^{S_{R}} \sum_{j=1}^{N_{k}} F_{i,k}} 
\end{equation}
After the simplification we have:
\begin{equation}
  P_{i,k} = \frac{\sum_{j=1}^{N_{k}} H(q^{ij}-q^{min})}{S_R M_{k}}   
\end{equation}
where $H(\alpha)$ is
\begin{equation*}
H(\alpha) = \\
\begin{cases}
1 & \text{if $\alpha > 0$}\\
0 & \text{otherwise} 
\end{cases}
\end{equation*}
which is the eq. 11 in the main text. $N_{k}$, $S_{R}$ and $M_{k}$ are the abundance of species $k$, the total number of extant species in community $J_{R}$ and the total number of potential mating interactions within the species $k$, respectively.

This genotype fitness model has the following same ingredient than the
neutral unified model: (1) individuals have the same probability
$1/J_R$ ($1/J_P$) to be chosen for death, and the following additions:
(1) individuals equally connected within their own species are
equivalent in fitness, i.e, the identity to a given species does not
confer per se fitness advantage. Fitness differences are then
considered only within each species; (2) there is a density
dependence in fitness across species, thus rare species have
relatively higher probabilities of reproduction in comparison to 
our unified basic neutral model described above. Note also that in this model may happen that the offspring of highly connected parents can inherit their connectance, thus increasing its reproductive probability. In the same way, individuals are chosen for death and reproduction in the second community. We have explored this model with the probability for death inversely proportional to the reproductive probability and the results remain qualitatively similar to the results with the same probability to be chosen for death. In the same way, individuals are chosen for death and reproduction in the second community $J_{P}$, thus the description of the model is equivalent.

\subsection{Phenotype Fitness Speciation Model}

Similar to the neutral and the genotype fitness model individuals have the same probability $1/J_R$ ($1/J_P$) to be chosen for death. Each individual $i$ of species $k$ is chosen for reproduction with probability proportional to their fitness, thus

\begin{equation}
  P_{i,k}^B = \frac{\sum_{j=1}^{J_{P}} m_{ij}}{M_k}   
\end{equation}
 where the sum until $J_{p}$ means the total number of interactions of individual $i$ with all individuals of community $J_P$. $M_{k}$ is the total number of mutualistic interactions of all the individuals of species $k$ with all the individuals in community $J_{P}$:
\begin{equation}
  M_{k} = \sum_{i = 1}^{N_{k}} \sum_{j = 1}^{J_{P}} m_{ij}   
\end{equation}
We now use the normalization factor across all species:
\begin{equation}
  {\cal{N_{B}}} = \sum_{i=1}^{S_{R}} \sum_{j=1}^{N_{k}} P_{i,k}^B   
\end{equation}
and the probability of individual $i$ of having a newborn is:
\begin{equation}
  P_{i,k} = \frac{P_{i,k}^B}{\cal{N_B}}  
\end{equation}
Finally, after the simplification we have
\begin{equation}
  P_{i,k} = \frac{\sum_{j=1}^{J_{P}} m_{ij}}{S_R M_{k}}   
\end{equation}
which is the eq. 12 in the main text. The sum until $J_{P}$ means the total number of interactions of individual $i$ with all the individuals of community $J_{P}$. $m_{ij}$ means that there is an interaction between individual $i$ and $j$. $S_{R}$ and $M_{k}$ are the total number of extant species in community $J_{R}$, and the total number of mutualistic interactions among all the individuals of species $k$ with all the individuals in community $J_{P}$, respectively. This phenotype fitness model has the same two ingredients to the genotype model but working at ecological level: (1) individuals equally connected within their own species are equivalent in fitness, i.e, the identity to a given species does not confer per se fitness advantage. Fitness differences are then considered only within each species, (2) there is a density dependence in fitness across species, thus rare species have higher probabilities of reproduction, and (3) contrary to the genotype model, where the offspring inherits a number of potential matings from its parents, we assume that each offspring in this model starts with one trophic interaction. In the same way, individuals are chosen for death and reproduction in the second community. Similar to the genotype fitness model we have explored this model with the probability for death inversely proportional to the reproductive probability. Results remain qualitatively similar to the results with the same probability to be chosen for death.

\subsection{Sampling transients and the Steady State}

Recent work has emphasized the importance of transient dynamics rather than long-term behavior in ecological systems \cite{Hastings:2004}. In the present study we sampled transients and the steady state for the number of generations to speciation, genetic-species diversity and species richness. This will allow us to determine if transients and the long-term behavior are similar under the neutral, the phenotype and the genotype fitness models. We thus consider both aspects of the dynamics. 

In all our simulations, we have assumed a fully symmetric case:
$m_{N^{k}_{R},N^{k}_{P}}$~=~$m_{N^{k}_{P},N^{k}_{R}}$ between plant and pollinator species, and across species within 
the plant and pollinator community, $m_{N^{k}_{R},N^{k^{'}}_{P}}$~=~$m_{N^{k}_{P},N^{k^{'}}_{R}}$~=~$m$, and percapita
birth and death rates have been assumed to be taken the same value across 
species. Furthermore, in all the replicates we have simulated the
condition $q^{min}$ $>$ ${Q_{R}}^{*}$
$({Q_{P}}^{*})$. Given $J_{R}$ and $J_{P}$ individuals in the initial
population, a generation is an update of $J_{R}$ and $J_{P}$ time
steps. 

We have explored a set of initial parameter values. Mutation rates ($\mu$ from 0.001 to 0.0001), a minimum genetic similarity value to the development of viable and fertile offspring, $q^{min}$, from 0.75 to 0.95 in the context of a mutualistic effectiveness m = 1, and obligate mutualism, $\omega$ = 0. For the specific parameter combination of $J_{R}$ = $J_{P}$ = $10^{3}$, $\mu$ = $10^{-4}$, $q^{min}$ = $0.9$, m = 1, and $\omega$ = 0 we run 100 replicates with $10^{4}$ generations each. The equilibrium value for each replicate for each community was closed to $Q_{R}$ ($Q_{P}$) $\sim$  ${Q_{R}}^{*}$ (${Q_{P}}^{*}$) $\sim$ 0.7 for the neutral model. Results for all the parameter combinations explored were qualitatively similar. 

\newpage

\bibliographystyle{plain}
\bibliography{MelianetalB}
\newpage

\newpage
\subsection{Table Legend of Appendix}

$\bullet$ {Table A1 shows the variables, parameters and acronyms used in the order that they appear in the main part of the ms.}

$\bullet$ {Table A2 shows the different levels and mechanisms that can be considered in a general ``Unified Neutral Model''. Two general mechanisms and three levels give 8 possible combinations. Note that we consider here genetic and mating as different levels. Genetic level used in the present study assumes that mutation rates are equal for forward and backward mutations and across loci. This means equal fitness among all individuals within each population at that level (i.e., $f_{i} \approx f_{j}$). Mating behavior is constrained by the minimum genetic similarity value for viable and fertile offspring ($q^{ij}$ $>$ $q^{min}$) and can be neutral as in the neutral scenario (i.e., $f_{i} \approx f_{j}$, thus all individuals within each species are equivalent) or driven by the number of genetically related matings of each individual (i.e., $f_{i} \ne f_{j}$, with explicit differences within each species). Neutrality at ecological level assumes equivalence and symmetry in the feeding behavior ($m_{ij} = m_{ji} = m$). This neutral feeding behavior assumes competitive and mutualistic symmetric interactions of all individuals and the same percapita effect of each pollinator on each plant and viceversa. We explore here three scenarios. The first scenario represents the unified neutral model where individuals have the same fitness across all levels. This implies that each individual has the same probability to death or have descendants during the evolution of the system. This scenario is represented in the three ``Neutral'' conditions in the Table A2. In the phenotype and genotype models each individual has a fitness value given by the trophic degree or number of potential matings with individuals in the second community or within the same population, respectively. Fitness is defined as the sum of the total number of $j$ individuals each individual $i$ interact or can mate with in each time step. The phenotype model explores the same genetic conditions than the neutral case but with different trophic interactions among individuals within each population (i.e, $d_{i} \ne d_{j}$, thus the $X$ in the continuum, Table A3). This generates differences in the reproductive probabilities (i.e., $f_{i} \ne f_{j}$). Finally, the genotype fitness model explores different mating conditions than the neutral and the phenotype models (i.e., $f_{i} \ne f_{j}$ which implies the evolution of $d_{i} \ne d_{j}$, thus the $X$ at both levels, Table A4). Note that the phenotype and the genotype models end up with the same conditions but the work in opposite directions. Finally, we keep the same percapita effectiveness at ecological level in the three scenarios (i.e., $m_{ij} = m_{ji} = m$).}

\subsection{Tables}
\vspace{0.25 in}

\begin{tabular}{l|c|c|}
\hline
 & {\bf Table A1} \\
\hline
{\bf Parameter} & {\bf Meaning} \\
\hline
{\bf $\theta_{b}$} & Biodiversity number, species level  \\
\hline
{\bf $J$} & Initial population size or community size  \\
\hline
{\bf $\mu$} & Mutation rate \\
\hline
{\bf $q^{min}$} & Genetic similarity constraint to have viable and fertile offspring \\
\hline
{\bf $NTME$} & Neutral Theory of Molecular Evolution \\
\hline
{\bf $NTB$} & Neutral Theory of Biodiversity \\
\hline
{\bf $\nu$} & Speciation rate \\
\hline
{\bf $NUM$} & Neutral Unified Model \\
\hline
{\bf $GF$} & Genotypic-fitness speciation model \\
\hline
{\bf $PF$} & Phenotypic-fitness speciation model \\
\hline
{\bf $J^{s}$} & Incipient species size \\
\hline
{\bf $q^{ij}$} & Genetic similarity between individual $i$ and $j$ \\
\hline
{\bf $Q = [q^{ij}]$} & Genetic similarity matrix \\
\hline
{\bf $RI$} & Reproductive isolation \\
\hline
{\bf $\theta_{m}$} & Diversity molecular level \\
\hline
{\bf $S_{e}$} & Effective number of species \\
\hline
{\bf $J_{R}$} & Initial population size or community size of resource/plant species \\
\hline
{\bf $J_{P}$} & Initial population size or community size of pollinator species \\
\hline
{\bf $J_{m}$} & Total number of individuals \\
\hline
{\bf $Q_{R}$ = $[q^{ij}_{R}]$} & Genetic similarity matrix for the plant community \\
\hline
{\bf $Q_{P}$ = $[q^{ij}_{P}]$} & Genetic similarity matrix for the pollinator community \\
\hline
\end{tabular}

\newpage
\begin{tabular}{l|c|c|c|c|}
\hline
{\bf Level/Mechanism} & {\bf Neutral} & \hspace{0.25 in} {\bf Niche} \\
\hline\hline
\hspace{-0.2 in} {\bf Genetic} ($\mu_{S_{u}^i} \approx \ne \mu$) & \hspace{-0.15 in} $f_{i} \approx f_{j}$ & \hspace{0.25 in} $f_{i} \ne f_{j}$ \hspace{0.1 in} \\
\hline
\hspace{-0.2 in} {\bf Mating} ($q^{ij}$ $>$ $q^{min}$) & \hspace{-0.15 in} $f_{i} \approx f_{j}$ & \hspace{0.25 in} $f_{i} \ne f_{j}$ \hspace{0.1 in} \\
\hline
\hspace{-0.2 in}  {\bf Feeding}  & \hspace{-0.1 in} $d_{i} \approx d_{j}$,$m_{ij} = m_{ji} = m$ & \hspace{0.25 in} $d_{i} \ne d_{j}$,$m_{ij} \ne m_{ji}$ \\
\hline
\end{tabular}
\center{Table A2}
\vspace{0.3 in}

\begin{tabular}{l|c|c|c|c|c|}
\hline
{\bf Level/Mechanism} & {\bf Neutral} & \hspace{0.25 in} {\bf Niche} & \hspace{0.25 in} {\bf Continuum} \\
\hline\hline
{\bf Genetic} ($\mu_{S_{u}^i} \approx \ne \mu$) & \hspace{0.25 in} $f_{i} \approx f_{j}$ & \hspace{1.25 in} & \\
\hline
{\bf Mating} ($q^{ij}$ $>$ $q^{min}$) & & \hspace{0.65 in} $f_{i} \ne f_{j}$ \hspace{0.5 in} & X \\
\hline
{\bf Feeding} & \hspace{-0.05 in} $m_{ij} = m_{ji} = m$ & \hspace{0.6 in} $d_{i} \ne d_{j}$ \hspace{0.5 in} & X \\
\hline
\end{tabular}
\center{Table A3: Phenotype Fitness Speciation Model}
\vspace{0.3 in}

\begin{tabular}{l|c|c|c|c|c|}
\hline
{\bf Level/Mechanism} & {\bf Neutral} & \hspace{0.25 in} {\bf Niche} & \hspace{0.25 in} {\bf Continuum} \\
\hline
{\bf Genetic} ($\mu_{S_{u}^i} \approx \ne \mu$) & \hspace{0.25 in} $f_{i} \approx f_{j}$ & \hspace{1.25 in} & \\
\hline
{\bf Mating} ($q^{ij}$ $>$ $q^{min}$) & & \hspace{0.65 in} $f_{i} \ne f_{j}$ \hspace{0.5 in} & X \\
\hline
{\bf Feeding} & \hspace{-0.05 in} $m_{ij} = m_{ji} = m$ & \hspace{0.6 in} $d_{i} \ne d_{j}$ \hspace{0.5 in} & X \\
\hline
\end{tabular}
\center{Table A4: Genotype Fitness Speciation Model}